\newif\ifjcp
\jcptrue  
 
\ifjcp
    \documentclass[twocolumn,aip,jcp,reprint,floatfix]{revtex4-1}
\else
    \documentclass[journal=jacsat,manuscript=article]{achemso}
    \setkeys{acs}{articletitle = true, doi = true}
    \newcommand{\onlinecite}[1]{\hspace{-1 ex} \nocite{#1}\citenum{#1}} 
\fi

\usepackage{float}
\usepackage{xcolor}
\usepackage{lipsum}  
\usepackage{acro}
\usepackage{amsmath}
\usepackage{bm, upgreek}
\usepackage{algorithm}
\usepackage{algorithmicx}
\usepackage{algpseudocode}
\usepackage{hyperref}
\usepackage[capitalise]{cleveref}
\usepackage{graphicx}
\usepackage{siunitx}
\usepackage{tabularx}

\DeclareAcronym{rpa}{short=RPA, long=random phase approximation}
\DeclareAcronym{tda}{short=TDA, long=Tamm--Dancoff approximation}
\DeclareAcronym{mpi}{short=MPI, long=message passing interface}
\DeclareAcronym{evgw}{short=ev$GW$, long=eigenvalue self-consistent $GW$}
\DeclareAcronym{scgw}{short=sc$GW$, long=self-consistent $GW$}
\DeclareAcronym{qsgw}{short=qs$GW$, long=quasiparticle self-consistent $GW$}
\DeclareAcronym{fsgw}{short=fs$GW$, long=Fock matrix self-consistent $GW$}
\DeclareAcronym{df}{short=DF, long=density fitting}
\DeclareAcronym{ip}{short=IP, long=ionisation potential}
\DeclareAcronym{ea}{short=EA, long=electron affinity, plural=electron affinities}
\DeclareAcronym{mse}{short=MSE, long=mean signed error}
\DeclareAcronym{mae}{short=MAE, long=mean absolute error}

\usepackage{soul}
\newcommand{\toadd}[1]{{#1}}
\newcommand{\toremove}[1]{{}}

\crefformat{equation}{#2Eq.~#1#3}
\crefformat{algorithm}{#2Alg.~#1#3}
\crefformat{table}{#2Table~#1#3}
\crefformat{figure}{#2Figure~#1#3}

\ifjcp
\newcommand{\setinfo}{%
    \title{Self-consistent $GW$ via conservation of spectral moments}%
    \author{Oliver J. Backhouse}
    \thanks{These authors contributed equally to this work and share first authorship.}
    \affiliation{Department of Physics, King's College London, Strand, London WC2R 2LS, U.K.}%
    \author{Marcus K. Allen}
    \thanks{These authors contributed equally to this work and share first authorship.}
    \affiliation{Department of Physics, King's College London, Strand, London WC2R 2LS, U.K.}%
    \author{Charles J. C. Scott}
    \affiliation{Department of Physics, King's College London, Strand, London WC2R 2LS, U.K.}%
    \author{George H. Booth}%
    \email{george.booth@kcl.ac.uk}%
    \affiliation{Department of Physics, King's College London, Strand, London WC2R 2LS, U.K.}%
}
\else
\newcommand{\setinfo}{%
    \title{Self-consistent $GW$ via conservation of spectral moments}%
    \author{Oliver J. Backhouse}
    \altaffiliation{These authors contributed equally to this work and share first authorship.}
    \affiliation{Department of Physics, King's College London, Strand, London WC2R 2LS, U.K.}%
    \author{Marcus K. Allen}
    \altaffiliation{These authors contributed equally to this work and share first authorship.}
    \affiliation{Department of Physics, King's College London, Strand, London WC2R 2LS, U.K.}%
    \author{Charles J. C. Scott}
    \affiliation{Department of Physics, King's College London, Strand, London WC2R 2LS, U.K.}%
    \author{George H. Booth}%
    \email{george.booth@kcl.ac.uk}%
    \affiliation{Department of Physics, King's College London, Strand, London WC2R 2LS, U.K.}%
}
\fi

\ifjcp
\else
\fi

\ifjcp\else
    \setinfo
    \SectionNumbersOn
\fi

\newcolumntype{?}{!{\vrule width 1pt}}
\newcolumntype{P}[1]{>{\centering\arraybackslash}p{#1}}

\begin{document}

\ifjcp
    \setinfo
\fi

\begin{abstract}
    We expand on a recently introduced alternate framework for $GW$ simulation of charged excitations [Scott et. al., J. Chem. Phys., {\bf 158}, 124102 (2023)], based around the conservation of directly computed spectral moments of the $GW$ self-energy. Featuring a number of desirable formal properties over other implementations, we also detail efficiency improvements and a parallelism strategy, resulting in an implementation with a demonstrable similar scaling to an established Hartree--Fock code, with only an order of magnitude increase in cost. We also detail the applicability of a range of self-consistent $GW$ variants within this framework, including a scheme for full self-consistency of all dynamical variables, whilst avoiding the Matsubara axis or analytic continuation, allowing formal convergence at zero temperature. By investigating a range of self-consistency protocols over the $GW$100 molecular test set, we find that a little-explored self-consistent variant based around a simpler coupled chemical potential and Fock matrix optimization to be the most accurate self-consistent $GW$ approach. Additionally, we validate recently observed evidence that Tamm--Dancoff based screening approximations within $GW$ lead to higher accuracy than traditional random phase approximation screening over these molecular test cases. Finally, we consider the Chlorophyll A molecule, finding agreement with experiment within the experimental uncertainty, and a description of the full-frequency spectrum of charged excitations.
\end{abstract}

\ifjcp
    \maketitle
\fi


\section{Introduction}

Green's function methodologies in electronic structure hold a particular appeal, owing to their compact description of correlated effects within this dynamical variable compared to many wave function approaches ~\cite{Marie2024}. Furthermore, this dynamical information directly encodes the spectrum of quasiparticle (charged) excitations, critical for understanding the reaction mechanisms, optoelectronic and transport properties of correlated systems, and provides a direct link to many experimental spectroscopic probes. The effect of electron correlation on Green's functions is generally formulated via a self-energy, which is an object with the same mathematical properties as a Green's function, and encodes these correlation driven changes to the Green's function. These self-energies can be diagrammatically constructed in a rigorous and systematic fashion via many-body perturbation theory, which leads to a range of approaches which can be categorized by their diagrammatic content.

Arguably the most widespread of these is the $GW$ approximation, which dresses the quasiparticles found from a starting mean-field picture with a self-energy composed of an infinite resummation of bubble diagrams~\cite{Marie2024, Reining2017, Golze2019}. Applied within a density functional theory starting point, this results in a cost-effective approach for correlation-driven changes of charged excitations. Its performance has been especially validated for weakly correlated semiconductors and metals where it describes the long-range collective plasmonic excitations which can often dominate the many-body physics in extended systems, with it also recently growing in popularity in molecular contexts~\cite{VanSetten2013}. This first-order perturbative expansion of the self-energy in terms of the dynamically screened Coulomb interaction ($W(\omega)$) can be derived via a neglect of the vertex function within Hedin's equations \cite{Hedin1965,Hedin1999} . Furthermore, corrections to other dynamical quantities such as the optical excitation spectrum often begins from a $GW$ picture of the quasiparticles~\cite{Monino2023}, placing $GW$ in an important and central position in the context of spectral perturbation theories.

Without any further numerical approximations, $GW$ formally scales as $\mathcal{O}[N^6]$ (where $N$ is a generic measure of system size in this context) and are therefore rare, since there are a multitude of well-founded, efficient and improvable numerical approximations which can reduce the scaling down to $\mathcal{O}[N^4]$\cite{Veril2018, QuinteroMonsebaiz2022, Bintrim2021, Pokhilko2021, Duchemin2020, Bruneval2012}, $\mathcal{O}[N^3]$ \cite{Foerster2011, Kresse2016, Ma2021, Duchemin2021, Yeh2023} or even linear scaling with additional assumptions concerning locality or stochastic averaging~\cite{Kutepov2020,Schurkus2016}. These have led to different variants of $GW$ implementations depending on these additional numerical steps and the domains in which the dynamical variables are represented. These include plasmon-pole, analytic continuation, contour-deformation, space-time and other variants \cite{Tolle2024b,Dongming2025}. These all have different strengths and weaknesses depending on the energy scales of interest, accuracy required or dominating physical processes, and while some approximations can be systematically reduced to convergence, others (such as analytic continuation steps) can not be, despite much progress~\cite{Tolle2024}. Other approximations which are commonly employed include the quasiparticle approximation to the Dyson equation, where only the diagonal part of the self-energy (in the molecular orbital representation) is considered, precluding any relaxation of the electron density due to the correlations, but allowing each molecular orbital to be relaxed independently. 

Recently, there have been interest in `full frequency' reformulations of $GW$ theory, where the dynamical self-energy is represented in a supermatrix or upfolded representation as explicit degrees of freedom, avoiding the need to consider numerical grids in which the self-energy or Green's functions are described~\cite{Tolle2023, Tolle2024b, Gao2024, Backhouse2022, Scott2023, Rebolini2016, Veril2020, QuinteroMonsebaiz2022, Bintrim2021}.
This entirely removes the need to consider numerical convergence of these grids, to resolve the dynamics and transformations between domains. 
These `static' upfolded representations can nevertheless return to dynamical variables as discrete pole representations on the physical real-frequency at the end of the calculation, avoiding the need for ill-conditioned analytic continuation techniques which can be required in other formulations~\cite{Gazizova2024, Yeh2022}. 
One conceptual viewpoint that can be useful is that these upfolded representations can be considered as exploiting the duality between the effect of a self-energy and a hybridization \cite{Sriluckshmy2021}. 
The upfolded self-energy can be thought of as hybridization with an external space, which mimics the effect of correlation on the mean-field quasiparticle states by introducing retardation effects on the electron and hole propagation once this external space is later traced out.

In contrast to an `exact' upfolded formulation of $GW$ where the dynamics of the self-energy is considered in its entirety~\cite{Tolle2023,Bintrim2021}, in Ref.~\onlinecite{Scott2023} some of the authors recently introduced a `moment-conserving' upfolded $GW$ approach. In this, the upfolded representation can be directly constructed so as to exactly describe only a desired number of (separate particle and hole) {\em spectral moments} of the real-frequency dynamical distribution of the self-energy. These moments can be alternatively considered as an effective short-time expansion of the particle and hole self-energy, and is also related to other orthogonal polynomial expansions for frequency-domain dynamical objects~\cite{Weibe2006}. This approach leads to a particularly compact and compressed form for the upfolded representation, with the mean-field \toadd{reference} Fock matrix coupled to this effective environment, with the size of this external environment scaling only linearly with the orbitals in the system and the number of self-energy spectral moments to conserve. 

We note that this is the correct physical scaling of the information content for a self-energy for large systems. A scaling in the number of explicit poles which is greater than linear would at some system size no longer be able to be resolved on any finite grid of frequency points, with poles merging under infinitesimal broadening. Therefore, this moment expansion represents a physical compression of the self-energy for realistic systems. We also found this representation to be efficiently computable at the $GW$ level, and rapidly converges to the full spectral dynamics with respect to this moment order over all energies, since it is not specifically targeting any particular (e.g. frontier) excitation energy scale. This compact and rapidly-convergent expansion in terms of spectral moments can also be used to formulate variants of other spectral methods, including GF2\cite{Backhouse2020a, Backhouse2020b, Backhouse2021} and (extended) dynamical mean-field theory\cite{Nusspickel2020, Fertitta2018, PhysRevB.104.245114}, as well as being a alternative perspective in which to motivate new approximations\cite{Scott2024}.

The upfolded moment expansion formulation of $GW$ has a number of desirable properties. The numerically challenging `frequency integration' in the convolution to obtain the self-energy can be computed exactly for a given number of moments, avoiding a diagonal representation of this quantity. This allows an effective application of the Dyson equation rather than the solution to a diagonal quasiparticle approximation, including low-weighted spectral features in the resulting Green's function, density relaxation and avoidance of the numerical uncertainties which can be associated with the solution to the underdetermined quasiparticle equation (see e.g. Ref.~\onlinecite{Schambeck2024}). Furthermore, the upfolded yet only linear in size effective Hamiltonian can be completely diagonalized, allowing for the entire real-frequency spectrum to be obtained in a single shot, without the need to define the self-energy or Green's functions on a numerical grid or analytic continuation.
These numerical approximations are not the only source of differences between $GW$ implementations. A myriad of self-consistent approximations have also been considered in the $GW$ literature, which can relax the Green's function and/or screened Coulomb interaction used to define the self-energy~\cite{Kotani2007, Stan2009, Pokhilko2024}. This can remove the undesirable starting point dependence of $GW$~\cite{Bruneval2013}, and has been shown to allow some more strongly correlated systems to be tackled in this framework~\cite{Pokhilko2022, Ammar2024}. 

We first review of the moment-conserving $GW$ formulation and detail technical improvements in Sec.~\ref{sec:momGW}, including large-scale parallelism, spin-integration, and use of natural auxiliary functions to improve the performance of the method, demonstrating its scaling with respect to both CPU and memory costs. In Sec.~\ref{sec:self_consistency}, we further demonstrate that all widely-considered self-consistent $GW$ variants (including full self-consistency) can be straightforwardly included within its scope. In Sec.~\ref{sec:gw100} we apply the approach across a widely-used molecular test set, where we demonstrate the convergence with respect to self-energy moments, and across the different self-consistent approximations. 

We also consider the use of interactions screened both at the \ac{rpa} level, as well as the \ac{tda} which neglects one of the time-orderings of the bubble diagrams in the screening channel~\cite{Bintrim2021}. We find that this TDA screening outperforms RPA in this molecular context, and results in a particularly simple computational moment expansion approach. Furthermore, in Sec.~\ref{sec:fsgw} we detail the implementation of a largely unexplored variant of self-consistency which we dub `Fock self-consistency', where the density matrix is made consistent across the mean-field and correlated levels of theory. In Sec.~\ref{sec:gw100}, we find this to be the most accurate variant across the considered test set while remaining highly efficient, which indicates significant potential in its use in molecular contexts. 
Finally, in Sec.~\ref{sec:chlorophyll} we apply the approach for the full frequency dynamics of the Chlorophyll molecule, with both TDA and RPA screening, and observe the convergence of the frontier excitations and quasiparticle renormalization with respect to moment order. With TDA screening, the result aligns well with an experimental result for the first ionization potential.

\section{Moment-conserved GW} \label{sec:momGW}

In this section we recap the moment-conserving $GW$ approach introduced in Ref.~\onlinecite{Scott2023}, as well as highlighting a number of algorithmic improvements, optimizations and parallelism that depart from the previous presentation. We also provide the formulation in terms of spatial orbitals, assuming a restricted, time-reversal symmetric reference state, in contrast to the previous spin-orbital expressions.
The central step of $GW$ is the construction of the dynamic self-energy from the convolution of the Green's function and screened Coulomb interaction as a first order perturbative expansion,
\begin{equation}
    \Sigma_{pq}(\omega)=\frac{i}{2\pi} \sum_{rs} \int d\omega' G_{rs}(\omega + \omega'+i0^+) W_{pr,qs} (\omega'+i0^+),
\end{equation}
where $p, q, \dots$ denote the canonical molecular orbitals of the reference state\toadd{, and $0^+$ is an infinitessimal positive regularization}.
However, we are only interested in obtaining \toadd{a truncated series of} {\em spectral moments} of the separated hole \toadd{($<$)} and particle \toadd{($>$)} self-energy, defined at order $n$ as
\begin{align}
\Sigma^{(n,<)}_{pq} &= -\frac{1}{\pi} \int_{-\infty}^{\mu} \mathrm{Im}[\Sigma(\omega)_{pq}] \omega^n d\omega \label{eq:lesser_moments} \\
&= (-1)^n 
 \left. \frac{d^n \Sigma(\tau)_{pq}}{d \tau^n} \right|_{\tau=0^+} ,
\end{align}
and similarly
\begin{align}
\Sigma^{(n,>)}_{pq} &= \frac{1}{\pi} \int_{\mu}^{\infty} \mathrm{Im}[\Sigma(\omega)_{pq}] \omega^n d\omega \label{eq:greater_moments}\\
&= (-1)^n \left. \frac{d^n \Sigma(\tau)_{pq}}{d \tau^n} \right|_{\tau=0^-} ,
\end{align}
where \toadd{$\tau$ is imaginary time and} $\mu$ represents the chemical potential of the system.
This exposes the relationship of \toadd{these moments of the dynamical distribution of the self-energy spectra} to a Taylor expansion of the short-time dynamics of the greater and lesser parts of the self-energy, with the moments defining the integrated weight, mean, variance, skew, and higher-order moments of the dynamical distribution of each element of the self-energy in the frequency domain. \toadd{In this work, this moment expansion is used as an alternate approach to truncate the resolution of the effective dynamical character of the self energy and provide a systematically improvable approximation to the fully dynamical and matrix-valued self-energy, without requiring explicit resolution on time or frequency grids, and allowing for manifestly zero temperature spectra.}

Additionally, this frequency integration can be performed exactly from knowledge of the corresponding spectral moments of $G(\omega)$ and $W(\omega)$, with the latter denoted as $W^{(n)}$ and defined by an analogous expression to the self-energy moments of Eqs.~\ref{eq:lesser_moments} and\toadd{~\ref{eq:greater_moments}}. Assuming a diagonal representation of $G(\omega)$, as would be provided by a mean-field state in the canonical basis of hole (denoted $i, j, k, \dots$ of dimensionality $o$) and particle (denoted $a, b, c, \dots$ of dimensionality $v$) states, then we can write the spectral moments of the occupied and virtual self-energy as
\begin{align}
    \label{eq:convolution_occ}
    \Sigma^{(n, <)}_{pq}
    &=
    \sum_{t=0}^{n}
    \sum_{k}
    \begin{pmatrix}
        n \\ t
    \end{pmatrix}
    \left( -1 \right)^{t}
    \epsilon_{k}^{n-t}
    W^{(n)}_{pk, qk}
    \\
    \label{eq:convolution_vir}
    \Sigma^{(n, >)}_{pq}
    &=
    \sum_{t=0}^{n}
    \sum_{c}
    \begin{pmatrix}
        n \\ t
    \end{pmatrix}
    \epsilon_{c}^{n-t}
    W^{(n)}_{pc, qc}
\end{align}
where $\epsilon_k$ and $\epsilon_c$ run over occupied and virtual state energies of the Green's function
respectively, and $n$ is the number of moments in the expansion. Our parallelized implementation distributes this contraction index ($k$ and $c$ respectively) over \ac{mpi} processes. The moments of the screened Coulomb interaction require knowledge of the neutral excitation spectrum via a density response function, obtained in this work either at the RPA or TDA level. The construction of the moments of the screened Coulomb interaction, $W^{(n)}$, will be described in Sec.~\ref{sec:Wmom}. 
In addition to the dynamic self-energy above, a static self-energy is required to remove the explicit contribution of the exchange-correlation functional and ensure a well-defined perturbative expansion. This is given by  
\begin{align}
    \Sigma_{\infty} 
    &=
    \mathbf{K}\toadd{[\mathbf{D}]}
    -
    \mathbf{V_{xc}} \label{eq:StaticSE}
\end{align}
where $\mathbf{K}$\toadd{$[\mathbf{D}]$} is the exchange matrix (constructed from the density \toadd{matrix, $\mathbf{D}$,} defined by $G(\omega)$) and $\mathbf{V_{xc}}$ is the exchange-correlation potential of the reference state. This cancels to zero for a canonical Hartree--Fock reference state.

Once the hole and particle spectral moments of $\Sigma^{(n, </>)}_{pq}$ are found, an effective Hamiltonian of dimension $\mathcal{O}[n_\mathrm{mom}^\mathrm{max} (o+v)]$ can be constructed, where $n_\mathrm{mom}^\mathrm{max}$ indicates the maximum conserved moment order, and whose eigenvalues and eigenvectors define the pole energies and residues of the resulting correlated $GW$ Green's function. This effective Hamiltonian can be constructed by inverting the block Lanczos three-term recurrence relation, and substituting in the spectral moments as powers of this effective hamiltonian, as shown in Ref.~\onlinecite{Scott2023}. This results in the form
\begin{equation}
\bm{\tilde{H}} = 
\begin{bmatrix}
\mathbf{f} + \mathbf{\Sigma}_{\infty} & \mathbf{\tilde{W}} \\
\mathbf{\tilde{W}}^\dagger & \mathbf{\tilde{d}}
\end{bmatrix} , \label{eq:effH}
\end{equation}
where $\mathbf{f}$ is the reference Fock matrix in the MO basis, \toadd{which for $G_0W_0$ is just a diagonal matrix of the $\epsilon_k$ and $\epsilon_c$ orbital energies. The matrices $\mathbf{\tilde{W}}$ and $\mathbf{\tilde{d}}$ define a coupling of the molecular `physical' orbitals to an `external' space, describing the correlation-driven changes to the quasiparticle spectrum. The dimensionality of this external space (and hence the matrices) grows only linearly with moment order and system size. These matrices can be obtained via a series of recursive linear algebra steps according to the block Lanczos algorithm, starting from the $\Sigma^{(n, <)}_{pq}$ and $\Sigma^{(n, >)}_{pq}$ moments from Eq.~\ref{eq:convolution_occ}-\ref{eq:convolution_vir}, and providing a block tridiagonal matrix representation of the upfolded self-energy which exactly conserves these moments (see Ref.~\onlinecite{Scott2023} for more details)}. This external space is therefore simply an upfolded representation of a dynamical self-energy, which approximates the exact $GW$ self-energy up to the moment truncation, and can be written in a `downfolded` form as
\begin{equation}
    \Sigma_{pq}(\omega) = \mathbf{\tilde{W}}^\dagger (\omega \mathbf{1}-\mathbf{\tilde{d}})^{-1} \mathbf{\tilde{W}} . \label{eq:downfoldedSE}
\end{equation}
\toadd{Since the dimensionality of $\mathbf{\tilde{d}}$ grows linearly with system size and moment order, we can see from Eq.~\ref{eq:downfoldedSE} that the number of poles of the effective self-energy also grows in this way.}
All dressed quasiparticle excitations are found from the complete diagonalization of this Hamiltonian, $\bm{\tilde{H}}$. In this way, the correlated $GW$ retarded Green's function can be subsequently obtained in a Lehmann representation in the frequency domain as
\begin{equation}
    G(\omega)_{pq} = \sum_\alpha \frac{\chi_{p \alpha} \chi_{q \alpha}^*}{\omega - E_\alpha + i \delta} , \label{eq:ret_GF}
\end{equation}
where $\chi_{p \alpha}$ are the eigenvectors of the effective Hamiltonian of Eq.~\ref{eq:effH}, with $p$ denoting an index in the MO basis of the physical space, and $\alpha$ label the eigenstates of the full system with eigenvalues $E_\alpha$, with $\delta$ a regularizing Lorentzian broadening for the spectral function. The spectral function can then be found as 
\begin{equation}
    A(\omega) = -\frac{1}{\pi} \sum_{p} \textrm{Im}[ G(\omega)_{pp}] . \label{eq:spectrum}
\end{equation}

Since the dimension of the effective Hamiltonian of Eq.~\ref{eq:effH} scales linearly with the number of moments, increasing the moment order correspondingly increases the number of poles in the resulting spectral function of Eq.~\ref{eq:spectrum}. \toadd{In this way, the moment order can be seen to directly control the effective resolution of the dynamics of the self-energy (and thus Green's function) across the entire spectral range, with increased splitting of the self-energy poles at higher moment orders.} Furthermore, since the numerator of the Green's function of Eq.~\ref{eq:ret_GF} corresponds to the component of the eigenvector corresponding to each excitation in the physical space (i.e. once the external space is traced out), this can be less than unity, describing a suppression of spectral weight due to the correlations and a reduced quasiparticle weight of the excitation in the spectrum.
This contrasts with more traditional $GW$ approaches where each excitation is obtained independently via a self-consistent solution (or linearized approximation) of the quasiparticle
equation, assuming a diagonal self-energy~\cite{Dongming2025}. \toadd{Allowing for the coupling between the bare quasiparticle states afforded by non-diagonal elements of the self-energy (moments) admits electron density relaxation by mixing these states (even at the $G_0W_0$ level), and can also (along with the dynamics) split the quasiparticle states in more strongly correlated systems~\cite{10.1063/5.0250929}. The coupled external space also results} in an increased number of solutions compared to the quasiparticle approximation, additionally providing lower-weighted \toadd{incoherent} excitations associated with satellite peaks~\cite{Loos2018b, Veril2018}.

\subsection{Screened Coulomb interaction moments and Natural auxiliary functions} \label{sec:Wmom}

The construction of the moments of the screened Coulomb interaction (required for Eqs.~\ref{eq:convolution_occ}-\ref{eq:convolution_vir}) can be achieved with a contraction of the moments of the reducible density response function with (static) Coulomb interaction. 
We first consider a low-rank decomposition of the Coulomb interaction as 
\begin{equation}
(ij|kl) = \sum_P^{N_\text{aux}} V_{P,ij} V_{P,kl} ,
\end{equation}
where $P, Q, \dots$ denotes an auxiliary basis of dimension $N_\mathrm{aux}$, and $(ij|kl)$ are the electron repulsion integrals in Mullikan (`chemists') notation. These low-rank decompositions can be achieved with standard approaches such as density fitting or Cholesky decompositions~\cite{Fukaya2014, Fukaya2020}. In this work, we further compress the dimensionality of the auxiliary basis into a basis of {\em natural} auxiliary functions (NAF)~\cite{Kallay2014}, of dimension $N_\textrm{NAF}$, as previously used in the context of $GW$ in Ref.~\onlinecite{Tolle2024}. These are constructed by forming and subsequently diagonalizing the positive semi-definite matrix
\begin{equation}
    M_{PQ}= \sum_{vw} V_{P,vw} V_{Q,vw}^*, \label{eq:NAF}
\end{equation}
whose eigenvectors define the NAF basis as contractions over the original auxiliary basis. This basis is truncated according to a threshold on the eigenvalues, $\epsilon_{\textrm{NAF}}$, which for an employed value of $\epsilon_{\textrm{NAF}}=10^{-5}$ was found to introduce negligible error into our results. The interaction tensor $\mathbf{V}$ is projected into this NAF basis, and can now be considered to have dimensions $N_\textrm{NAF} \times M^2$, where $M=o+v$ is the total number of orbitals. The result of this compression is that $N_\textrm{NAF} < N_\textrm{aux}$, resulting in a speed up of the most computationally demanding step of the full algorithm by a factor $(N_\textrm{aux}/N_\textrm{NAF})^2$. For representative truncations with conservative thresholds, this can result in time savings of $\sim0.75^2$, e.g. for the largest alkane considered in Fig.~\ref{fig:benchmark} ($C_{32}H_{66}$).


In principle, the orbital pair $(v,w)$ should run over the entire $M^2$ space to construct the natural auxiliaries which best approximate the entire interaction tensor in a least-squares sense. However, by default in this work we do not contract over the virtual-virtual block of the interaction, which we find leads to only a minor differences in compactness but reduces the cost of implementing Eq.~\ref{eq:NAF} (as also found in Ref.~\onlinecite{Tolle2024}). \toadd{This insensitivity to the virtual-virtual block in the efficiency of the NAF compression is driven by the fact that the occupied-virtual interaction terms are the dominant in $GW$ theory, defining the interaction kernel of the RPA, while the virtual-virtual block is only used in the subsequent screening equation (see Eq.~\ref{eq:w_moments} below).} The exception to this is self-consistent implementations where the screened Coulomb interaction itself is updated (see Sec.~\ref{sec:self_consistency}). In this, we find increased sensitivity to the choice of orbital spaces contracted over in Eq.~\ref{eq:NAF} and therefore include the full orbital product space in the contraction. This is due to the fact that the NAF basis \toadd{itself is not reconstructed through the self-consistency in our implementation, and therefore should be constructed in a way which is insensitive to the definition of the precise choice of orbital spaces}. Full self-consistency starts to mix virtual-virtual interaction channels of the original mean-field state into the screening as the propagators are updated, and therefore these channels also need to be accurately described in the initial construction of the NAF basis.


Once the compressed low-rank interaction tensors have been constructed, the moments of the screened Coulomb interaction are given as
\begin{align}
    \label{eq:w_moments}
    W_{px, qx}^{(n)}
    &=
    \sum_{PQ}
    \sum_{ia}
    V^*_{P, px}
    \tilde{\eta}^{(n)}_{P, ia}
    V^*_{Q, ia}
    V_{Q, qx},
\end{align}
where $x$ enumerates the hole and particle states in the Green's function, but where only the hole {\em or} particle states of $x$ are required for definition of the lesser (Eq.~\ref{eq:convolution_occ}) and greater (Eq.~\ref{eq:convolution_vir}) parts of the self-energy moments respectively. The tensors $\mathbf{W}^{(n)}$ are built with the $x$ index distributed over MPI ranks, while the contraction over $(i,a)$ is also distributed and reduced over these processes. The quantity $\tilde{\eta}^{(n)}_{P,ia}$ is derived from the reducible density response \toadd{for the neutral excitation spectrum,} defined by the screening model, which can be \toadd{directly constructed} either at the RPA or TDA level as we will describe next. 

\subsection{RPA screening} \label{sec:rpa}

We can define the spectral moments of the reducible density response function, $\eta^{(n)}_{ia,jb}$, providing the appropriate screening either at the \ac{rpa} or \ac{tda} level. \toadd{We can connect these moments to the traditional dynamical derivation of $GW$ theory, by first defining the dynamical density response function of RPA as} 
\begin{equation}
    \chi(\omega)_{pq,rs} = (\mathbf{P}(\omega)^{-1}-\mathcal{K})^{-1} ,
\end{equation}
\toadd{where $\mathbf{P}(\omega)$ is the irreducible polarizability defined by the reference state, and $\mathcal{K}$ is the interaction kernel over the particle-hole channel which couples the bubble diagrams. We can sum over the particle-hole excitations and de-excitations, as}
\begin{equation}
    \eta(\omega)_{ia,jb} = \chi(\omega)_{ia,jb} + \chi(\omega)_{ia,bj} + \chi(\omega)_{ai,jb} + \chi(\omega)_{ai,bj} .
\end{equation}
\toadd{We can then integrate over frequency in order to define the moments of the dynamical distribution of $\eta(\omega)$ for a systematically improvable expansion, as}
\begin{equation}
\eta^{(n)}_{ia,jb} = -\frac{1}{\pi}\int_0^{\infty} \text{Im}[\eta(\omega+i0^+)_{ia,jb}] \omega^n d\omega .
\end{equation}
We can furthermore define the useful intermediate, $\tilde{\eta}^{(n)}_{P,ia}$, as
\begin{align}
    \label{eq:alt_dd}
    \tilde{\eta}^{(n)}_{P, jb} = \sum_{ia} V_{P, ia} \eta^{(n)}_{ia,jb}.
\end{align}

\toadd{While the above connects the $\tilde{\eta}^{(n)}$ quantity to the dynamical formulation of the density response in the RPA, in practice we can avoid the need to directly perform this frequency integration for each moment, and instead efficiently construct these moments via a recursive algorithm, motivated in Ref.~\onlinecite{Scott2023} from the equation-of-motion formulation of the RPA. To do this, we first} define a diagonal matrix $\mathbf{D}$ of the energy differences of the reference state (poles of the irreducible polarizability) as
\begin{align}
    \label{eq:energy_differences}
    D_{ia,jb}
    =
    (\epsilon_{a} - \epsilon_{i}) \delta_{ab}\delta_{ij} ,
\end{align}
as well the matrices corresponding to the blocks of the standard equation-of-motion formulation of RPA theory~\cite{Heselmann2010, Chen2017} as
\begin{align}
    \mathbf{A}
    &=
    \mathbf{D}
    +
    2\mathbf{V}^{\dagger} \mathbf{V}
    ,
    \\
    \mathbf{B}
    &=
    2\mathbf{V}^{\dagger} \mathbf{V}
    .
\end{align}
We note that all interaction tensors in \ac{rpa} (and \ac{tda}) are defined over the particle-hole $(i,a)$ channel, meaning $\mathbf{V}=V_{P,ia}$ for the construction of these density function moments, which we have previously constructed in the NAF basis and distributed over \ac{mpi} ranks (Sec.~\ref{sec:Wmom}). \toadd{The additional factor of $2$ is added to account for spin integration over these restricted spatial orbitals.}
The zeroth moment of the reducible density response function at the level of \ac{rpa} can then be calculated as
\begin{align}
    \tilde{\bm{\upeta}}^{(0)}
    &=
    \mathbf{G}
    \mathbf{D}^{-1}
    \left(
        \mathbf{I}
        -4\mathbf{V}^{\dagger}\left(
        \mathbf{I}+4\mathbf{V}\mathbf{D}^{-1}\mathbf{V}^{\dagger}\right)^{-1}
        \mathbf{V}\mathbf{D}^{-1}\right).
\end{align}
%
The $\mathbf{G}$ intermediate is found efficiently via numerical integration, related to implementations of total energy calculations for \ac{rpa} correlation energies~\cite{Furche2008, Eshuis2010}. It is broken into a sum over two parts, $\mathbf{G}_{\mathrm{main}}$ and $\mathbf{G}_{\mathrm{offset}}$, both of dimensionality $N_\textrm{NAF} \times ov$, defined as
\begin{align}
    \label{eq:Gmain}
    \mathbf{G}_{\mathrm{main}}
    &=
    \frac{1}{\pi}
    \int_{-\infty}^{\infty}
    z^{2}
    \mathbf{Q} (z)
    \left(
        \left(
            \mathbf{I}
            +
            \mathbf{Q} (z)
        \right)^{-1}
        -
        \mathbf{I}
    \right)
    \mathbf{V}
    \mathbf{F} (z)
    dz
    ,
    \\
    \mathbf{G}_{\mathrm{offset}}
    &=
    \mathbf{V} \mathbf{D}
    +
    4
    \int_{0}^{\infty}
    \mathbf{V} \mathbf{D}
    \mathbf{E} (z)
    \mathbf{V}^{\dagger}
    \mathbf{V}
    \mathbf{E} (z)
    dz
    ,
\end{align}
with
\begin{align}
    \mathbf{Q} (z)
    &=
    4
    \mathbf{V}
    \mathbf{F} (z)
    \mathbf{D}
    \mathbf{V}^{\dagger}
    ,
    \\
    \mathbf{F} (z)
    &=
    \left(
        \mathbf{D}^{2} + z^{2}\mathbf{I}
    \right)^{-1}
    ,
    \\
    \mathbf{E} (z)
    &=
    \exp{
        \left(
            -z \mathbf{D}
        \right)
    }
    .
\end{align}
The numerical integration of $\mathbf{G}_{\mathrm{main}}$ is performed by Clenshaw–Curtis quadrature, with grid points and weights optimized for improved efficiency. This optimization and resulting exponential convergence of both integrals is described in detail in Ref.~\onlinecite{Scott2023}, with 24 integration points used for the largest system in this work (the Chlorophyll system of Sec.~\ref{sec:chlorophyll}) for a tight convergence of the integral.

The first order moment can be simply evaluated more simply, as
\begin{align}
    \tilde{\bm{\upeta}}^{(1)}
    &=
    \mathbf{V} \mathbf{D}
    ,
\end{align}
Higher order moments can then be found recursively from the knowledge of $\tilde{\bm{\upeta}}^{(0)}$ and $\tilde{\bm{\upeta}}^{(1)}$, as
\begin{align}
    \tilde{\bm{\upeta}}^{(n)}
    &=
    \tilde{\bm{\upeta}}^{(n-2)}
    \left(
        \mathbf{D}^{2}+4
        \mathbf{V}^{\dagger}
        \mathbf{V}\mathbf{D}
    \right).
\end{align}
A full derivation of these equations can be found in Ref.~\onlinecite{Scott2023}, with any changes to the final working expressions due to the spin-integrated formulation shown here, along with some efficiency improvements in Eq.~\ref{eq:Gmain} resulting from the minor modification to the definition of $\tilde{\bm{\upeta}}^{(n)}$ in Eq.~\ref{eq:alt_dd} of this work.

The distributed \ac{mpi} parallelism for the efficient evaluation of these density response moments, $\tilde{\bm{\upeta}}^{(n)}$, proceeds via distributing the $(i,a)$ compound index of the diagonal matrix, $\mathbf{D}$, which can be contracted with the correspondingly distributed $\mathbf{V}$ over these orbital index pairs. In contractions over this compound index, the result is then reduced across the \ac{mpi} processes. This leads to an efficient distributed-memory parallelism, and an overall computational scaling of $\mathcal{O}[N_\textrm{NAF}^2 o v + N_\textrm{NAF}^3]$.


\subsection{Tamm--Dancoff screening} \label{sec:tda}

A less diagrammatically complete form of screening can be found by the Tamm--Dancoff approximation (TDA). This considers screening of the interaction by the series of only forwards-in-time bubble diagrams, precluding a mixing of de-excitations in the construction of the density response. Nevertheless, the \ac{tda} approximation has shown indication in the literature that the resulting $GW$ spectrum can be more accurate in molecular systems~\cite{Bintrim2021}. Algebraically, this imposes the condition $\mathbf{B} = \mathbf{0}$, which significantly simplifies the moments of the density response in the \ac{tda}, which become
\begin{equation}
\bm{\upeta}^{(n)} = \mathbf{A}^{n}.
\end{equation}
The required computational variables $\tilde{\bm{\upeta}}^{(n)}$ can then be evaluated in $\mathcal{O}[N_\textrm{NAF}^2 ov]$ scaling, without any numerical integration \toadd{since $\mathbf{\eta}^{(0)} = \mathbf{I}$ within the TDA}.
The zeroth order moment within the \ac{tda} is therefore
\begin{align}
    \tilde{\bm{\upeta}}^{(0)}
    &=
    \mathbf{V}
    ,
\end{align}
while subsequent moments can be found according to the recurrence relation
\begin{align}
    \tilde{\bm{\upeta}}^{(n)}
    &=
    \tilde{\bm{\upeta}}^{(n-1)}
    \left(
        \mathbf{D}
        +
        2\mathbf{V}^{\dagger}
        \mathbf{V}
    \right)
    .
\end{align}
These expressions can be similarly parallelized over $(i,a)$, with significantly fewer floating point operations, albeit with the same formal scaling as the \ac{rpa} screening. We will also consider the use of \ac{tda} screening on the results and how this affects self-consistent implementations in Sec.~\ref{sec:resultsgw100}.

\subsection{Computational scaling benchmarks}

\begin{figure}
    \centering
    \includegraphics[width=0.49\textwidth]{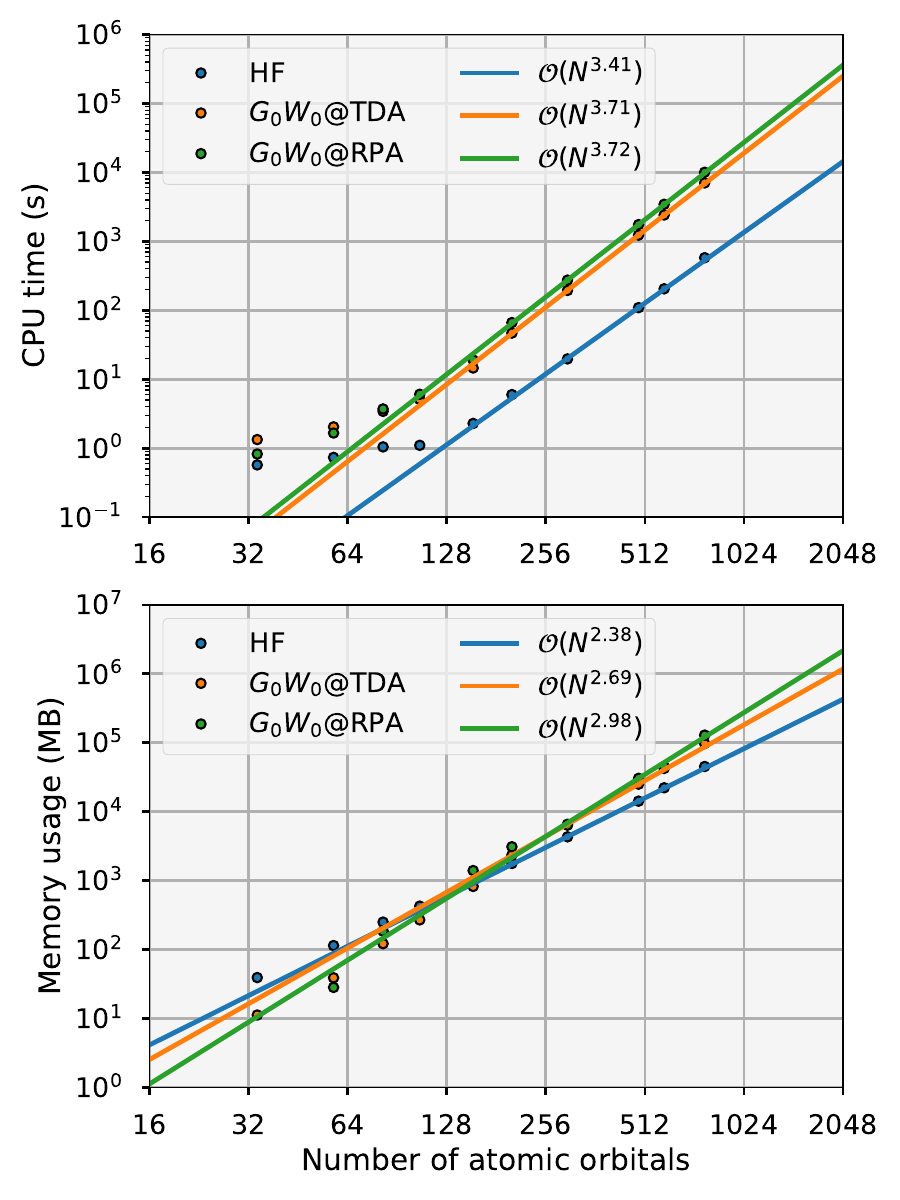}
    \caption{
        Performance benchmark for the moment-conserving $G_0W_0$ implementation, indicating CPU time (top) and memory usage (bottom) for
        increasing numbers of atomic orbitals in linear alkane chains up to $C_{32}H_{66}$ in a cc-pVDZ basis, with maximum moment order $n_{\mathrm{mom}}^{\mathrm{max}}=7$, obtaining the full $G_0W_0$ spectrum for each calculation. No natural auxiliary orbital compression was used in this consideration of this resource scaling.
        Lines are shown for both \ac{tda} and \ac{rpa} screening,
        and for Hartree--Fock via {\tt PySCF} for comparison.
    }
    \label{fig:benchmark}
    \ignorespacesafterend
\end{figure}

\cref{fig:benchmark} shows a benchmark in computational efficiency for our $GW$ procedure,
indicating the CPU time and memory usage
of the moment-conserving $G_0W_0$ method for
increasing numbers of atomic orbitals. We show results for both \ac{tda} and \ac{rpa} screening, and a fixed maximum moment order $n_{\mathrm{mom}}^{\mathrm{max}}=7$. The systems used for these calculations were increasing sizes of simple chain alkanes in a cc-pVDZ basis set, although at no point do we exploit any locality present in these quasi-one-dimensional systems to skew the observed scaling. The observed computational complexity for both the CPU time and memory usage for each method is shown in the respective legends via a fit to the largest systems. These asymptotic scalings are as expected, with all methods exhibiting roughly a quartic scaling
in their CPU time with system size, and a cubic scaling in memory usage. The cost of our $GW$ implementation is therefore only just over an order of magnitude more expensive than the default Hartree--Fock code in {\tt PySCF}~\cite{PySCF2017,PySCF2020}. The cubic-scaling memory cost is however currently the bottleneck for the code. While this can be partially alleviated through the usage of the NAFs, discussed in Sec.~\ref{sec:Wmom}, which would reduce memory cost by a factor of $\sim0.75$ and computational cost by a factor of nearly half for the largest alkane chain (though was not considered in the resource data of Fig.~\ref{fig:benchmark}), this memory bottleneck remains a priority area for future development, via the use of tensor hypercontraction or interpolative separable density fitting techniques~\cite{Yeh2023, Yeh2024, Pokhilko2024b, Lu2017}.

\section{Self-consistency} \label{sec:self_consistency}


One of the significant drawbacks of the most widespread single-shot `$G_0W_0$' is the dependence of the result on the choice of reference state, which in turn defines the initial energy levels for the irreducible polarizability ($\mathbf{D}$ in Eq.~\ref{eq:energy_differences}) and space of particle-hole excitations for the screening channel. While free from formal double-counting of correlated effects from the DFT and the $GW$ diagrammatic expansion, this maintains an undesirable dependence on the choice of exchange-correlation functional inherited from DFT and can make it more difficult to compare $GW$ results. Furthermore, it is unclear how to systematically improve the spectrum beyond $GW$, with proposals to increase the perturbative order in the expansion in terms of the screened Coulomb interaction, or including vertex corrections in the screening often met with mixed results\cite{Berkelbach2019, Forster2022, Bruneval2024}.

An alternative approach to alleviate some of these drawbacks is to define a self-consistent procedure on one or more of the variables in $GW$ that define the self-energy, ranging from \toadd{the reference state} or density-based self-consistency, to the dynamical Green's function or screened Coulomb interaction. Full self-consistency of all objects results in a formal conservation of desired expectation values within $GW$, but was initially found to yield poor results for the uniform electron gas~\cite{vonBarth1996}, although has recently found renewed interest~\cite{Harsha2024,Ren2021,Bruneval2021b}. A number of more heuristic self-consistencies have also been formulated in this field (which may not be formally conserving approximations), but which reduce or eliminate the dependence on the choice of exchange-correlation functional in the reference state. We find that essentially all of these approaches, from partial to full self-consistency, can also be straightforwardly implemented within this moment-based approach to $GW$, which we will describe in this section. We will then compare and contrast these self-consistent moment-truncated variants in Sec.~\ref{sec:resultsgw100}, comparing their convergence with maximum moment order and benchmarking the quality of results compared to high-accuracy reference data in molecular systems.

\subsection{One-shot $G_{0}W_{0}$} \label{sec:gw}

We first consider the one-shot $G_0W_0$ method \cite{Hybertsen1985} within the moment-conserving framework from an algorithmic perspective as a primitive routine in the self-consistent adaptations. All approaches assume that the compact natural auxiliary functions described in Sec.~\ref{sec:Wmom} have been computed in advance (if desired), and are used for the expansion of the factorized bare Coulomb interaction as input to the algorithm.
In the one-shot approach the moments of the self-energy are computed according to Eqs.~\ref{eq:convolution_occ}-\ref{eq:convolution_vir},
after which they are used to find the effective Hamiltonian of Eq.~\ref{eq:effH}. The diagonalization of this Hamiltonian provides the set of poles of both the self-energy and correlated Green's function as shown in Eq.~\ref{eq:ret_GF}. Given the moments, the construction and diagonalization of this upfolded moment-conserving Hamiltonian is achieved with the \texttt{MBLSE} solver within the
\texttt{dyson} package. 



\begin{algorithm}[H]
    \caption{$G_{0}W_{0}$ implementation in \texttt{momentGW}}
    \label{alg:g0w0}
    \begin{algorithmic}
        \Function{$G_0W_0$}{
            $\bm{\upepsilon}$,
            $\mathbf{C}$,
            $\mathbf{V}^{\mathrm{AO}}$,
            $n_{\mathrm{mom}}^{\mathrm{max}}$
        }
            \State $f_{p,p}, \Sigma_{\infty,pq} \gets \epsilon_{p} \delta_{pq}$ \Comment{\cref{eq:StaticSE}}
            \State $V_{Q, ia} \gets \sum_{\alpha \beta} V^{\mathrm{AO}}_{Q, \alpha \beta} C_{\alpha i}^{*} C_{\beta a}$ \Comment{Distribute over $(i,a)$}
            \State $\hat{V}_{Q, px} \gets \sum_{\alpha \beta}  V^{\mathrm{AO}}_{Q, \alpha \beta} C_{\alpha p}^* C_{\beta x}$ \Comment{Distribute over $x$}
            \State $\mathbf{D} \gets \bm{\upepsilon}$
            \Comment{\cref{eq:energy_differences}}
            \State $\tilde{\bm{\upeta}}^{(n)} \gets \mathbf{V}, \mathbf{D}$
            \Comment{\cref{sec:rpa} or \cref{sec:tda}}
            \State $\mathbf{W}^{(n)} \gets \mathbf{V}, \hat{\mathbf{V}}, \tilde{\bm{\upeta}}^{(n)}$
            \Comment{\cref{eq:w_moments}}
            \State $\bm{\Sigma}^{(n, <)} \gets \bm{\upepsilon}^{G}, \mathbf{W}^{(n)}$
            \Comment{\cref{eq:convolution_occ}}
            \State $\bm{\Sigma}^{(n, >)} \gets \bm{\upepsilon}^{G}, \mathbf{W}^{(n)}$
            \Comment{\cref{eq:convolution_vir}}
            \State $\mathbf{\tilde{H}} \gets f_{p,p}, \bm{\Sigma}_\infty, \bm{\Sigma}^{(n, <)}, \bm{\Sigma}^{(n, >)}$
            \Comment{\cref{eq:effH}, \texttt{dyson.MBLSE}}
            \State $\bm{G}(\omega) \gets \mathbf{\tilde{H}}$ \Comment{\cref{eq:ret_GF}} 
        \EndFunction
    \end{algorithmic}
\end{algorithm}

The overall $G_0W_0$ algorithm is sketched schematically in \cref{alg:g0w0}, whose input is a set of orbital energies from a DFT or Hartree--Fock reference, $\bm{\upepsilon}$, the transformation from atomic (AO) to molecular orbitals (MO), $\mathbf{C}$, the factorized bare Coulomb interaction in the AO and NAF basis, $\mathbf{V}^{\mathrm{AO}}$, and the maximum order to which the self-energy moments will be computed, $n_{\mathrm{mom}}^{\mathrm{max}}$. Note that in the algorithm, two copies of the three-index Coulomb interaction are transformed, in order to facilitate the parallelism of the algorithm. The first, $V_{Q, ia}$, only contains the interaction in the particle-hole channel, and is distributed over MPI threads by the compound index $(i,a)$, while the second which we denote $\hat{V}_{Q, px}$, spans the entire MO product space, and is distributed over MPI threads via the $x$ index which labels the index contracted with the Green's function in the final convolution with the screened Coulomb moments (Eq.~\ref{eq:w_moments}) in Eq.~\ref{eq:convolution_occ}-\ref{eq:convolution_vir}.

\subsection{\Acl{evgw}} \label{sec:evgw}

Eigenvalue self-consistent $GW$ (ev$GW$) performs a partial self-consistent solution to $GW$
where the orbital energies are updated between iterations without updating the orbital coefficients or overall electron densities\cite{Faleev2004, Shishkin2007, Blase2011a, Kaplan2015}. 
This is particularly convenient as the orbital space remains the same dimension, the bare Coulomb interaction tensors remain unchanged, and the existing expressions can be modified by simply updating $\bm{\upepsilon}$. However, since the orbitals do not change, it does not \toadd{fully} remove the reference dependence of the $GW$ starting point. There is an additional choice as to whether it is either the reference Green's function, the screened Coulomb interaction or both that is modified through the iterations to reflect the updated orbital energies in the reference state. If only the Green's function is changed (denoted ev$GW_0$), then only the $\bm{\upepsilon}$ values in Eqs.~\ref{eq:convolution_occ}-\ref{eq:convolution_vir} are updated each iteration, while if the screened Coulomb interaction is modified to reflect changes in the irreducible polarizability moments, then Eq.~\ref{eq:energy_differences} is also changed and the screened Coulomb moments must be regenerated each iteration. Where both $G$ and $W$ moments are updated, we will denote the approach ev$GW$.

In order to update the quasi-particle energies each iteration within this scheme, it is necessary to associate a single distinct eigenvalue of the effective Hamiltonian, $E_\alpha$, with each MO. We do this by associating the pole of the correlated Green's function indexed by $\alpha$ with the largest quasiparticle weight on each MO index $p$, as determined by the eigenvectors, $\chi_{p \alpha}$, by using
\begin{align}
    \label{eq:quasiparticle_energy_update}
    \epsilon_{p}^{\mathrm{QP}}
    &=
    E_{\alpha}
    \ \
    \mathrm{where}
    \ \
    \alpha
    =
    \arg\max_{\beta}
    \left\vert \chi_{p \beta} \right\vert^2 .
\end{align}
For systems where each $GW$ excitation can be simply associated with a quasi-particle of the original reference state, this condition should be easily fulfilled, and therefore neglects the presence of satellite peaks in the spectrum for the update of the quasiparticle energies. A schematic of the implemented \ac{evgw} algorithm can be found in Appendix~\ref{app:algos}. 

\subsection{\Acl{qsgw}} \label{sec:qsgw}

\Acf{qsgw} achieves self-consistency by casting the self-energy to a static potential, $\mathbf{V}^{\Sigma}$, and using this potential to update the self-consistent field which defines the reference state, its single-particle energies and coefficients \cite{VanSchilfgaarde2006, Kotani2007b, Kutepov2012, Koval2014}. The self-consistent field is reoptimized as
\begin{align}
    \label{eq:qp_scf}
    \left(
        \mathbf{f} [\mathbf{C}]
        +
        \mathbf{V}^{\Sigma}
    \right)
    \mathbf{C}
    &=
    \mathbf{C}
    \bm{\upepsilon}
    ,
\end{align}
where $\mathbf{C}, \bm{\upepsilon}$ are an updated set of single particle orbitals and energies, which can be used to calculate a new $GW$ self-energy, followed by a new static potential, and so forth until convergence.
In the real-frequency formulation of qs$GW$, this effective static potential simply evaluates the self-energy at the previous iteration quasi-particle energies, with a finite broadening $\delta$, as
\begin{align}
    \label{eq:qsgw_v}
    V_{pq}^{\Sigma}
    &=
    \frac{1}{2}
    \mathrm{Re}
    \left[
        \Sigma_{pq} (\epsilon_{p})
        +
        \Sigma_{pq} (\epsilon_{q})
    \right]
    .
\end{align} 

The self-energy evaluated at these specific frequencies can be found by directly considering the $\tilde{W}$ and $\tilde{d}$ blocks of the effective Hamiltonian, $\bm{\tilde{H}}$ in Eq.~\ref{eq:downfoldedSE}. By diagonalizing $\tilde{d}$ and rotating ${\tilde{W}}$ to a representation where the external degrees of freedom have been decoupled, a set of couplings between the MO and external spaces can be found, $v_{p \lambda}$, along with the energies of the external degrees of freedom, $\varepsilon_\lambda$. This allows for the effective dynamical self-energy to be computed as a sum over discrete poles as
\begin{align}
    \label{eq:qsgw_se_eta}
    \Sigma_{pq} (\epsilon_{p})
    &=
    \sum_{\lambda}
    \frac{
        v_{p\lambda} v_{q\lambda}^{*}
    }{
        \epsilon_{p}
        -
        \varepsilon_{\lambda}
        -
        \mathrm{sign}(\varepsilon_{\lambda})
        i\delta
    }
    .
\end{align}
%
This method of regularisation of the self-energy on the real-frequency with a Lorentzian broadening can present problems, and results were found to vary with choice of $\delta$.
In response to this, the similarity renormalisation group (SRG) \cite{Monino2022,Marie2023} approach to regularisation in \ac{qsgw} calculations was developed in Ref.~\onlinecite{Marie2023} which uses 
\begin{align}
    \label{eq:qsgw_se_srg}
    \Sigma_{pq} (\epsilon_{p})
    &=
    \frac{
        \Delta_{p\lambda}
        +
        \Delta_{q\lambda}
    }{
        \Delta_{p\lambda}^{2}
        +
        \Delta_{q\lambda}^{2}
    }
    v_{p\lambda}
    v_{q\lambda}^{*}
    \left(
        1
        -
        \exp{
            \left(
                -[
                    \Delta_{p\lambda}^{2}
                    +
                    \Delta_{q\lambda}^{2}
                ]
                s
            \right)
        }
    \right)
    ,
\end{align}
with
\begin{align}
    \label{eq:qsgw_se_srg_delta}
    \Delta_{p\lambda}
    &=
    \epsilon_{p}
    -
    \varepsilon_{\lambda}
    .
\end{align}
This procedure was \toadd{found to allow for smoother convergence and reduced sensitivity with respect to the regularization parameter, $s$~\cite{Marie2023}. Both} flavors of \ac{qsgw} are available in our \texttt{momentGW} code. 

\toadd{An even more recent variant, which still appeals to a self-consistent quasi-particle approximation, but which avoids the need to define a static, Hermitian potential from the self-energy, was introduced by Duchemin and Blase in Ref.~\onlinecite{10.1063/5.0250929}. In this, the diagonality of the Green's function at the quasi-particle energies is maximized via a numerical optimization of a unitary rotation of the orbital basis. These orbitals are then used to define the particle-hole excitation channel for the screening.}

We note that at convergence of qs$GW$ approaches, there are two Green's functions which can be considered; there is the correlated Green's function obtained in the same fashion as the previous methods (Eq.~\ref{eq:ret_GF}), while a Green's function can also be constructed from the effective self-consistent mean-field description of Eq.~\ref{eq:qp_scf}. As is traditional in the qs$GW$ literature we choose the latter, where all excitations have a unit quasi-particle weight, and no satellites can be included in the description~\cite{Kotani2007}.
Our implementation of this moment-based qs$GW$ scheme is detailed further in Appendix~\ref{app:algos}.

\subsection{\Acl{scgw}} \label{sec:scgw}

\Acf{scgw} offers more complete self-consistency \cite{Holm1998, Stan2006, Caruso2012, Grumet2018, Koval2014}, fully updating the Green's function and (optionally) screened Coulomb interaction to construct them directly from the {\em correlated} Green's function of the previous iteration, as given in Eq.~\ref{eq:ret_GF}. This not only changes the orbital coefficients, but also increases the {\em number} of excitations in the `reference' state defining the irreducible polarization propagator, each with different weight, avoiding the requirement of assigning `quasi-particle-like' excitations as required in ev$GW$ (Sec.~\ref{sec:evgw}). This results in a rigorously `conserving' $GW$ approximation, \toadd{which ensures that appropriate physical quantities such as spin, charge, energy and momentum are conserved, gauge freedoms are retained, along with consistency between thermodynamic relationships as functional derivatives. From a practical perspective, this also ensures a} reference independent solution to the Dyson equation in the full-frequency limit. To the best of our knowledge, fully self-consistent $GW$ has always been performed on the Matsubara axis at finite temperature, to allow for the increasing structure of the optimized Green's function with iteration, accounting for the introduction of low-weighted satellite peaks into the self-consistent Green's function. However, in this work we show that the moment truncation allows for this self-consistency to formally be considered at zero temperature, and without an explicit grid representation of the Green's function, with the approach ensuring that the number of poles of the Green's function is naturally truncated each iteration via the moment conservation criterion.

However, the number of excitations in this self-consistent variant does grow larger than the original quasi-particles defined by the MO space, and this requires additionally rotating $\mathbf{V}$ and $\hat{\mathbf{V}}$ at each iteration, with an enlarged space for the compound $(i,a)$ index and $x$ index respectively in these objects. These indices now label all particle and hole eigenstates found in the diagonalization of the effective Hamiltonian $\bm{\tilde{H}}$ of Eq.~\ref{eq:effH} in each iteration, which we term `quasi-molecular orbitals' (QMOs).
In similar fashion to other self-consistent schemes, 
one can also define $GW_{0}$ by fixing the screened Coulomb interaction moments and keeping the $i,a$ indices referring to the original MO space,
or indeed $G_{0}W$ by fixing the space indexed by $x$ in Eq.~\ref{eq:w_moments} and the $\bm{\upepsilon}$ tensor in  Eqs.~\ref{eq:convolution_occ} and \ref{eq:convolution_vir}. 
The moment-conserving \ac{scgw} algorithm can also be found schematically in Appendix~\ref{app:algos}.

\subsection{\toadd{Electron number conservation}}

\toadd{Each time a new self-energy is constructed in an sc$GW$ step, it is necessary to find an appropriate chemical potential for the Green's function in order to conserve total electron number. Within our moment expansion approach, each pole of the Green's function is associated with an eigenvector of the constructed effective Hamiltonian. The component of this eigenvector in the physical space indicates the (generally non-integer) number of electrons which this pole contributes to the total electron number, i.e. its quasiparticle weight. Explicitly, once the effective Hamiltonian $\bm{\tilde{H}}$ has been found at the end of a $GW$ iteration (which includes the upfolded representation of the $GW$ self-energy), the trace of the correlated density matrix, $\rho_{pq}$, commensurate with the Green's function of Eq.~\ref{eq:ret_GF} can be found as a sum over the occupied eigenstates of $\bm{\tilde{H}}$, to give the physical electron number as
\begin{equation}
    N_\mathrm{elec} = \mathrm{Tr}[\rho_{pq}] = \mathrm{Tr} \left[ 2 \sum_{\alpha: E_\alpha < \mu} \chi_{p \alpha} \chi_{q \alpha} \right], \label{eq:density}
\end{equation}
where $\mu$ is the chemical potential of the system, ensuring that the sum runs only over occupied poles (eigenstates), $\alpha$, and where $p,q$ denote orbitals in the physical sector of the Hamiltonian. To define the electron number, it is therefore simply necessary to define which eigenvectors of $\bm{\tilde{H}}$ should be considered `occupied'. This reduces the `optimization' of a chemical potential to an Aufbau-like filling of the energy levels of the correlated Green's function, until the filling optimally satisfies the total electron number of Eq.~\ref{eq:density}. This avoids the need for iterative root-finding algorithms and repeated transforms for chemical potential optimizations in Matsubara formulations of sc$GW$.}

\toadd{In Sec.~\ref{sec:scgw}, it was argued that sc$GW$ should rigorously conserve electron number, however, this is not true through the intermediate iterations, nor for other approximations, and nor strictly away from the full-frequency limit (i.e. our necessarily truncated moment order). Therefore, we can find that the electron number in general will not be exactly conserved in the Green's function, i.e. there will {\em not} be an Aufbau filling of the discrete Green's function poles which results in the exact integer electron number. We note that this problem would also formally manifest in more traditional sc$GW$ implementations, where a chemical potential is optimized over the Matsubara axis. However, the finite grid resolution there implies a finite temperature, which smears out electron number, and means that finding a unique chemical potential which exactly satisfies electron number is possible. However, this is formally an artifact of the necessarily finite temperature which we do not have within the Lehmann representation of the Green's function in our algorithm.}

\toadd{In our moment-sc$GW$ approach, we ignore these small electron number errors, noting that the error decreases with increasing moment order, as increasing numbers of low-weighted poles emerge that enable the total electron number to be satisfied. However we can alternatively devise a strategy to explicitly modify the self-energy to {\em exactly} conserve electron number, even at finite moment order and zero temperature. This can be achieved by finding a small {\em relative} chemical potential offset between the self-energy and the static potential defining the reference state (i.e. $\mathbf{f}+\mathbf{\Sigma_\infty}$). This shifts the self-energy poles a small amount relative to $G_0(\omega)$, which in turn changes the correlated Green's function to enable the total electron number to be satisfied. In practice for moment-sc$GW$, this amounts to optimizing a small shift to the diagonal of the ${\mathbf{\tilde{d}}}$ matrix in Eq.~\ref{eq:effH} (i.e. the energy levels of the `external' space), until the total electron number of Eq.~\ref{eq:density} is satisfied to within any desired threshold at every iteration. However, while we believe that this should not result in any change to the final Green's function for sc$GW$ in the infinite moment limit, at finite moment order this resulted in a materially different correlated Green's function and increased the average IP over the $GW$100 test set. This relative self-energy shift to impose electron number was therefore not used in our sc$GW$ algorithm, but rather used as motivation for a new self-consistent $GW$ varient, when combined with a full self-consistent field optimization of the reference potential ($\mathbf{f}+\mathbf{\Sigma_\infty}$) each iteration, which is described further below.}

\subsection{\Acl{fsgw}} \label{sec:fsgw}

\toadd{In this section, we propose a modified self-consistent $GW$ algorithm, which combines the explicit electron number imposition described above, along with a self-consistent field optimization over the correlated density matrix each iteration. This can serve to relax the single-particle density in the presence of the effective dynamical self-energy, such that the Fock matrix (which defines $G_0(\omega)$ each iteration, along with the orbitals and their energies used to build the next $\Sigma(\omega)$) is consistent with the {\em correlated} density matrix. At convergence, this ensures a matching of the density matrix from the correlated Green's function to the density matrix obtained from the reference Fock matrix used to define the orbitals and energies which define the irreducible polarizability and hence self-energy. In addition, the electron number is explicitly enforced for the correlated Green's function at every step.}

\toadd{To achieve this, once a self-energy and correlated Green's function has been built, the density matrix resulting from this Green's function is first constructed via Eq.~\ref{eq:density}. This density matrix is used to form a new Fock matrix, $\mathcal{F}$, which will form the static component of the self-energy, represented in the top-left block of $\mathbf{\tilde{H}}$. There are then two nested self-consistent loops. In the first, a relative chemical potential offset, $\zeta$, is optimized between the (fixed) self-energy and $\mathcal{F}$, in order to ensure the correct electron number of the correlated Green's function. Once this has converged, $\mathcal{F}$ itself is optimized via a Hartree--Fock self-consistent field procedure, with the correlated (non-idempotent) density matrix at each step obtained from the full Green's function (i.e. in the presence of the fixed $\Sigma(\omega)$) until it is consistent with the density matrix produced by the diagonalization of the resulting single-particle Fock matrix. This chemical potential optimization and Fock matrix optimization are iterated in the presence of this fixed self-energy, until both are consistently satisfied, as the density matrix optimization can also cause the physical electron number to change. At this point, a new self-energy is constructed, using the orbitals and energies obtained from the optimized Fock matrix.}

This can be implemented straightforwardly within the moment-constrained approach, by manipulating the effective upfolded Hamiltonian of Eq.~\ref{eq:effH}, $\bm{\tilde{H}}$, found at the end of each $GW$ iteration. Its eigenbasis can define the correlated density matrix via Eq.~\ref{eq:density}. This density can in turn be \toadd{used to define a new Fock matrix, $\mathcal{F}[\rho]$, to update the static part of the upfolded Hamiltonian in $\bm{\tilde{H}}$, initially defined by the reference state $\mathbf{f}+\mathbf{\Sigma}_\infty$ at the start of the calculation.} 
For the conservation of electron number, the energies of the self-energy poles are shifted slightly by a constant $(\zeta)$ in order to minimize an objective function given by the electron number error. This relative chemical potential is easily introduced within the effective Hamiltonian as a shift applied to the $\bm{\tilde{d}}$ block of Eq.~\ref{eq:effH}, until Eq.~\ref{eq:density} is correct.
Diagonalizing $\bm{\tilde{H}}$ again following this shift can in turn result in a new correlated single-particle density matrix, which can be self-consistently optimized to convergence (with DIIS acceleration)~\cite{Pulay1980, Pulay1982}. Once both the density and electron number are converged, this resulting converged Fock matrix can then be used to define the molecular orbitals and energies for $G$ and $W$ in the construction of an updated self-energy for the next iteration, until the self-energy, density matrix and electron number are all converged and consistent.
This is exactly akin to the so-called Fock loop introduced in previous work on the
AGF2 method~\cite{Backhouse2020a, Backhouse2020b}. 

Ultimately, these Fock loops and chemical potential optimizations ensure that the electron number is exactly conserved and relaxes the single-particle density such that the resulting Fock matrix is a fixed point derived from the correlated $GW$ density. Since this can fully update the single-particle density matrix and resulting reference state, it should also formally remove the reference dependence of the $GW$ results. Note that this procedure would not be possible with a diagonal approximation to the self-energy.
\toadd{We note that beyond relaxing the orbitals and energies from their mean-field reference, an important feature is that it treats incoherent low-weighted excitations in the same way as the quasi-particle-like peaks in the self-consistency. This avoids the need to define quasi-particle-like excitations of the correlated Green's function for the self-consistency, which can become ill-defined in the presence of stronger correlation effects or even introduce discontinuities across a smoothly changing reaction coordinate. Nevertheless, this self-consistency is only defined at the level of the single-particle density matrix, and therefore the Green's function of the reference state and the correlated Green's function will still not match at convergence, unlike sc$GW$. However, this `Fock self-consistency' (fs$GW$) is a cheaper variant than the full dynamical self-consistency of \ac{scgw}, since the dimensionality of the physical space does not change. The algorithm for this self-consistent fs$GW$ variant is sketched in \cref{alg:fsgw}.}


\toadd{Related self-consistent $GW$ methods which rely on an optimization of the static density matrix have increasingly been considered in recent years, with a linearized formulation found to improve total energies and other dynamical properties~\cite{Bruneval2021,PhysRevB.108.125107}. Furthermore, the recent $\gamma$s$GW$ scheme introduced by Duchemin and Blase advocated for a similar self-consistency on the correlated density matrix, which was found to improve the ionization potentials over the $GW$100 test set~\cite{10.1063/5.0250929}. In this scheme, the correlated density matrix was used in order to define a self-consistent Fock matrix (in turn defining $G_0(\omega)$) and static component of the self-energy, similar to above. However, the orbitals and associated energies derived from this Fock matrix were not used to define the subsequent polarizability and dynamical self-energy, which instead was derived from an extraction of quasi-particle energies from the correlated Green's function. This contrasts with the energies and orbitals used in the dynamical self-energy also being derived directly from the self-consistent Fock matrix in the fs$GW$ approach presented here, as well as the relative chemical potential which we impose to enforce electron number conservation.}

\begin{algorithm}[H]
    \begin{algorithmic}
        \Function{fsGW}{
            $\bm{\upepsilon}$,
            $\mathbf{C}$,
            $\mathbf{V}^{\mathrm{AO}}$,
            $n_{\mathrm{mom}}^{\mathrm{max}}$
        }
            \State $\mathcal{F}_{pq} \gets \epsilon_{p}\delta_{pq} + \Sigma_{pq}(\infty) $ \Comment{\cref{eq:StaticSE}}
            \While{$\bm{\Sigma}^{(n)}$ \textbf{not} converged}
                \State $\tilde{d}, \tilde{W}, \mathcal{F} \gets$ \Call{GW}{
                    $\bm{\upepsilon}$,
                    $\mathbf{C}$,
                    $\mathbf{V}^{\mathrm{AO}}$,
                    $n_{\mathrm{mom}}^{\mathrm{max}}$
                }
                \Comment{\cref{alg:g0w0}} 
                \While{\Call{get\_elec\_num}{$\tilde{d}$, $\tilde{W}$, $\mathcal{F}$, $\zeta$} $\neq$ N}
                \State optimize $\zeta$ s.t. \Call{get\_elec\_num}{$\tilde{d}$, $\tilde{W}$, $\mathcal{F}$, $\zeta$}$=N$
                    \While{$\bm{\rho}$ \textbf{not} converged}
                        \State $\bm{\tilde{H}}(\zeta) \gets \tilde{d} + \zeta \bm{I}, \tilde{W}, \mathcal{F}$ \Comment{Eq.~\ref{eq:effH}}
                        \State $\bm{\chi}(\zeta) \mathbf{E}(\zeta) \bm{\chi}(\zeta)^\dagger \gets \bm{\tilde{H}}(\zeta)$
                        \State Choose best $\mu$ via Aufbau
                        \State $\bm{\rho} \gets \bm{\chi}(\zeta), \bm{E}(\zeta), \mu$ \Comment{Eq.~\ref{eq:density}}
                        \State $\mathcal{F}_{pq}[\bm{\rho}] \gets \bm{\rho}$
                    \EndWhile
                \EndWhile
            \State $\bm{\upepsilon}, \bm{C} \gets \mathcal{F}$
           \EndWhile
            \State $\bm{\tilde{H}}(\zeta) \gets \tilde{d} + \zeta \bm{I}, \tilde{W}, \mathcal{F}$ \Comment{Eq.~\ref{eq:effH}}
            \State $\bm{G}(\omega) \gets \mathbf{\tilde{H}}$ \Comment{\cref{eq:ret_GF}}
            %

        \EndFunction
        \\\hrulefill
        \Function{get\_elec\_num}{$\tilde{d}$, $\tilde{W}$, $\mathcal{F}$, $\zeta$}
            \State $\bm{\tilde{H}}(\zeta) \gets \tilde{d} + \zeta \bm{I}, \tilde{W}, \mathcal{F}$ \Comment{Eq.~\ref{eq:effH}}
            \State $\bm{\chi}(\zeta) \mathbf{E}(\zeta) \bm{\chi}(\zeta)^\dagger \gets \bm{\tilde{H}}(\zeta)$
            \State Choose best $\mu$ via Aufbau
            \State $\bm{\rho} \gets \bm{\chi}(\zeta), \bm{E}(\zeta), \mu$ \Comment{Eq.~\ref{eq:density}}
            \State \Return{$\textrm{Tr}[\bm{\rho}]$}
        \EndFunction
    \end{algorithmic}
    \caption{\Acf{fsgw} algorithm}
    \label{alg:fsgw}
\end{algorithm}

\section{$GW$100 Benchmark} \label{sec:gw100}

The proliferation of self-consistency variants (including optimization of $G$, $W$ or both), the choice of mean-field reference state, TDA or RPA screening as well as the property under consideration (e.g. frontier excitations or full spectrum) mean that it is hard to benchmark the full space of relevant quantities across $GW$ approaches. While several common technical considerations are removed in the approaches of this work (e.g. approximations to the convolution, analytic continuation or quasi-particle equation), there is also the question of convergence with the order of conserved moments, and how this affects the accuracy of these different variants. 
Within molecular $GW$ methods, the $GW$100 test set has emerged as a common benchmark for different implementations, enabling an assessment of approximations between numerical schemes and uncovering \toadd{deficiencies in certain design choices in $GW$ methodologies (such as} the approach used to solve the quasiparticle equation, which we avoid in all $GW$ variants in this work)~\cite{VanSetten2015}. This test set consists of a collection of 102 diverse molecules with a wide range of properties and bonding types. Note that while the test set was originally 100 systems, two systems have since received updates to their structures and so the set actually consists of 102 structures. This makes a suitable setting in which to benchmark the convergence and accuracy of the self-consistent moment-conserving $GW$ approaches of this work. 

\subsection{Exemplar molecule: Borane}
\label{sec:example}

\ifjcp
\begin{figure*}
    \centering
    \includegraphics[width=\textwidth]{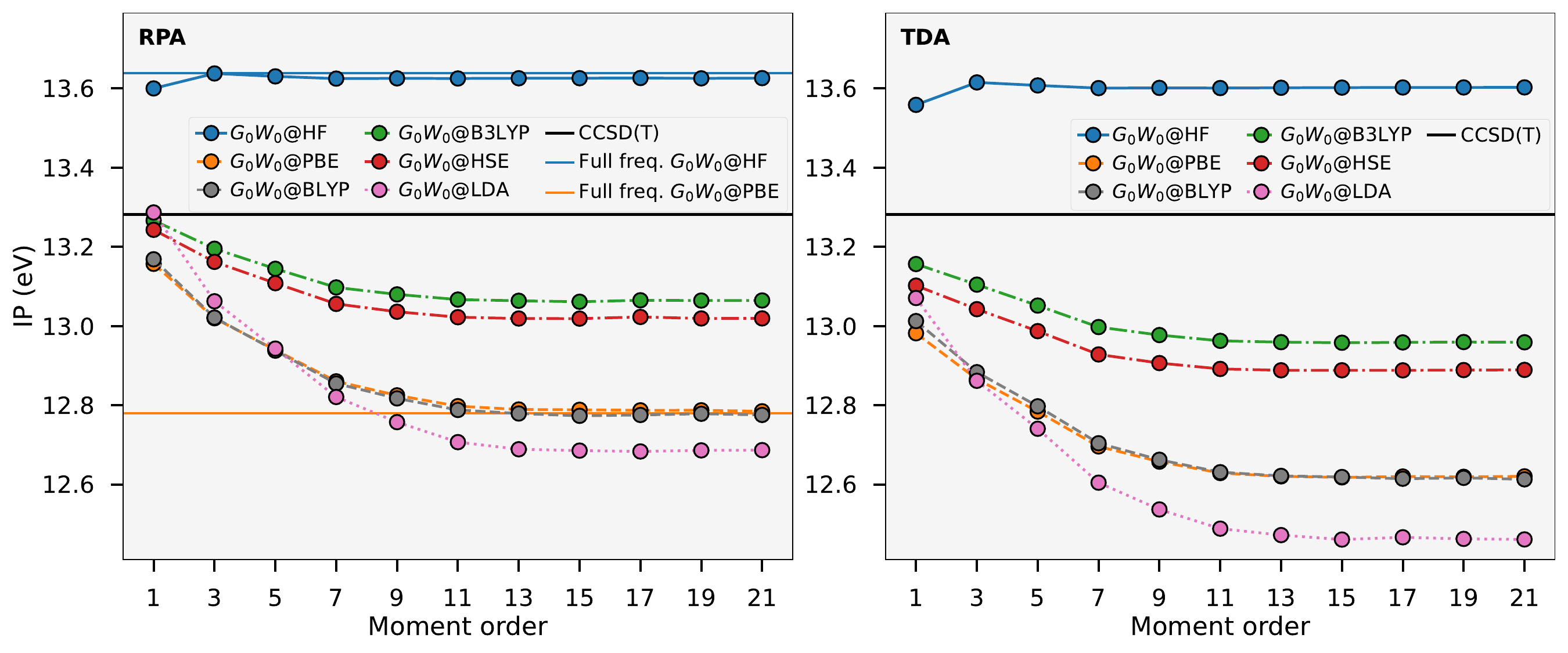}
    \caption{\toadd{Convergence of the IP of Borane ($BH_3$) with respect to the number of conserved moments ($n_\textrm{mom}^\textrm{max}$) of the self-energy in a def2-TZVPP basis set for single-shot $G_0W_0$, with RPA screening (left) and TDA (right). A range of mean-field starting points are considered, as well as reference values for RPA screening from {\tt PySCF}, implementing an $\mathcal{O}[N^6]$ full-frequency algorithm to remove any grid approximations \cite{doi:10.1021/acs.jctc.0c00704}. The remaining discrepancy likely comes from the diagonal approximation to the self-energy enforced in the reference values.}}
    \label{fig:borane_mf}
\end{figure*}

\begin{figure*}
    \centering
    \includegraphics[width=\textwidth]{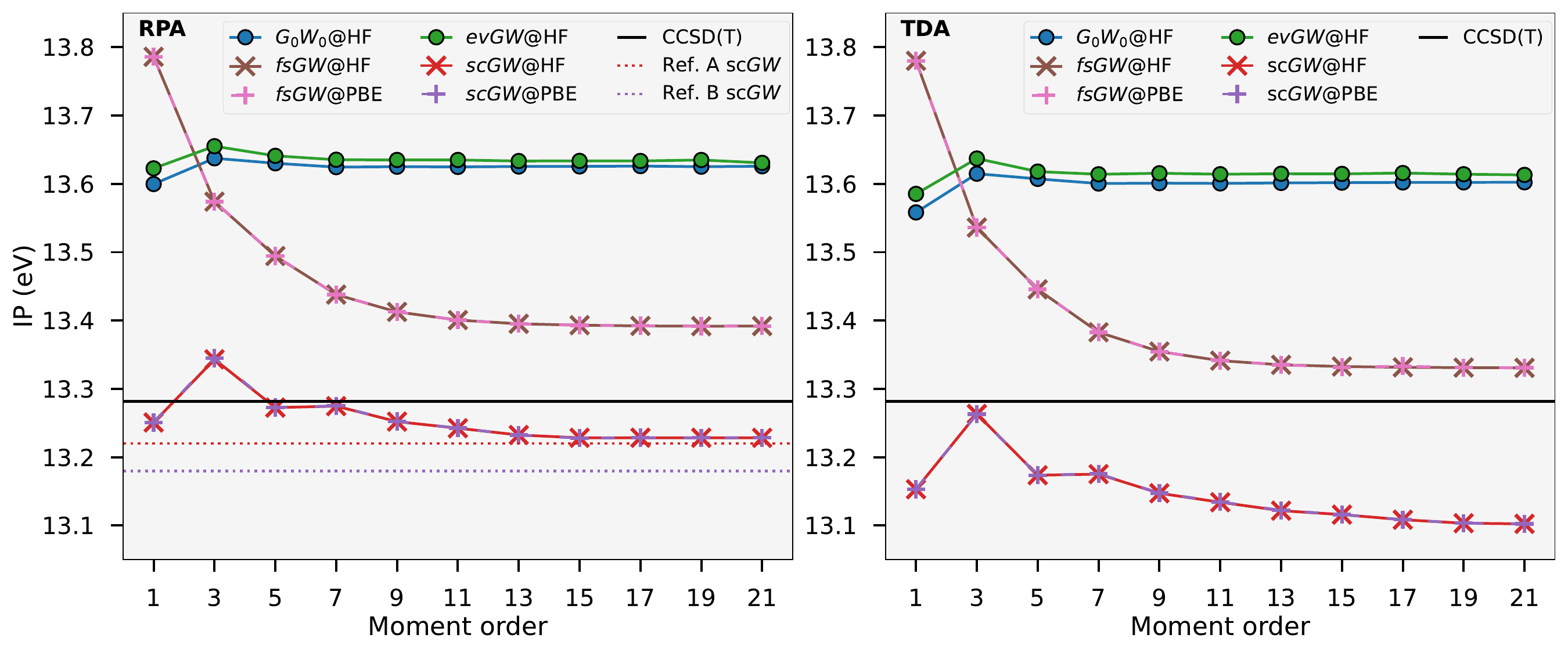}
    \caption{\toadd{Convergence of the IP of borane ($BH_3$) with respect to the number of conserved moments ($n_\textrm{mom}^\textrm{max}$) for self-consistent implementations of moment conserved $GW$ across HF and PBE reference states. We consider RPA (left) and TDA (right) screening, in a def2-TZVPP basis. Reference fully self-consistent sc$GW$ values are included from Caruso et al.~\cite{Caruso2016} (Ref. A) and Wen et al.~\cite{Wen2024} (Ref. B), relying on self-consistency on the Matsubara axis followed by analytic continuation.}}
    \label{fig:borane_gw}
\end{figure*}
\else
\begin{figure}
    \centering
    \includegraphics[width=\textwidth]{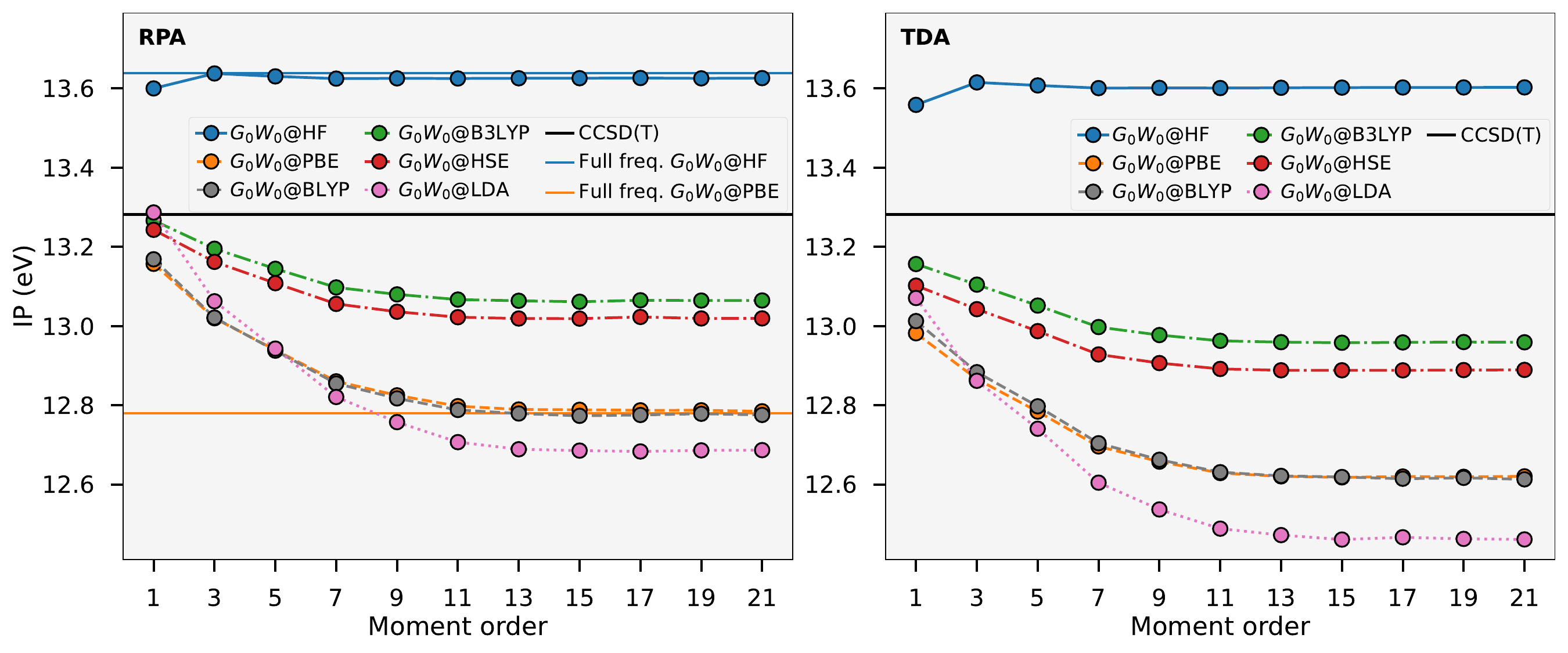}
    \caption{\toadd{Convergence of the IP of Borane (BH$_3$) with respect to the number of conserved moments ($n_\textrm{mom}^\textrm{max}$) of the self-energy in a def2-TZVPP basis set for single-shot $G_0W_0$, with RPA screening (left) and TDA (right). A range of mean-field starting points are considered, as well as reference values for RPA screening from {\tt PySCF}, implementing an $\mathcal{O}[N^6]$ full-frequency algorithm to remove any grid approximations \cite{doi:10.1021/acs.jctc.0c00704}. The remaining discrepancy likely comes from the diagonal approximation to the self-energy enforced in the reference values.}}
    \label{fig:borane_mf}
\end{figure}

\begin{figure}
    \centering
    \includegraphics[width=\textwidth]{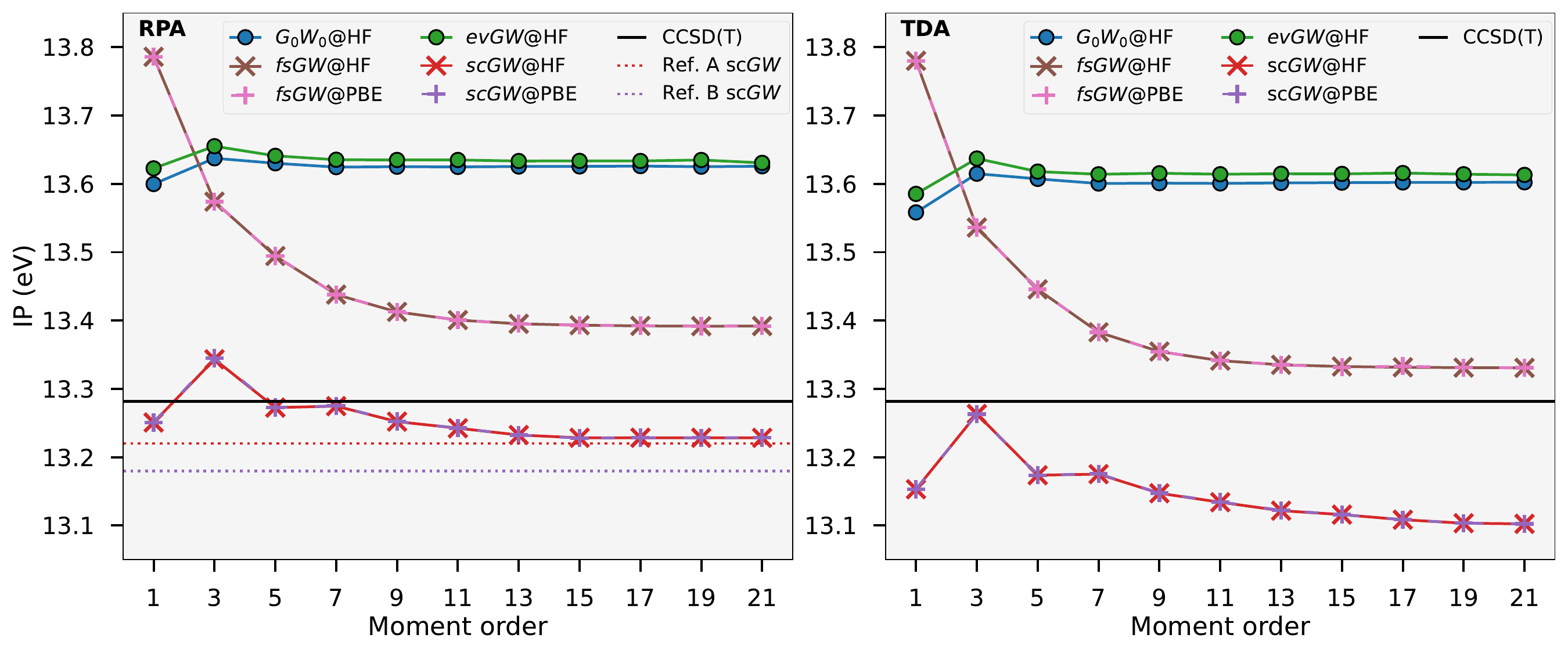}
    \caption{\toadd{Convergence of the IP of Borane (BH$_3$) with respect to the number of conserved moments ($n_\textrm{mom}^\textrm{max}$) for self-consistent implementations of moment conserved $GW$ across HF and PBE reference states. We consider RPA (left) and TDA (right) screening, in a def2-TZVPP basis. Reference fully self-consistent sc$GW$ values are included from Caruso et al.~\cite{Caruso2016} (Ref. A) and Wen et al.~\cite{Wen2024} (Ref. B), relying on self-consistency on the Matsubara axis followed by analytic continuation.}}
    \label{fig:borane_gw}
\end{figure}
\fi

\toadd{Before considering the aggregated results over the full test set, in Fig.~\ref{fig:borane_mf} we focus on the IP of Borane (BH$_3$) as an example, and to quantify the impact of mean-field starting point and moment order convergence in our $G_0W_0$ approach. A range of functionals are considered; two GGA functionals (PBE and BYLP), two hybrid functionals (HSE06 and B3LYP), a local density approximation (LDA) and Hartree--Fock (HF). It has been observed that allowing for the density relaxation effect afforded by removing the diagonal approximation for the self-energy (as we do in this work) can ameliorate some of this starting point dependence~\cite{10.1063/5.0250929}. In addition, we consider both RPA and TDA screening approximations. For the RPA screening, we compare to two reference ($\mathcal{O}[N^6]$ scaling) `full-frequency' results on HF and PBE references~\cite{doi:10.1021/acs.jctc.0c00704}, which remove uncertainties due to choices of numerical grids. We find good convergence to these reference values as the moment order increases, noting that the small remaining discrepancies (7meV for PBE and -13meV for HF references) likely arise from the diagonal approximation to the self-energy in these reference values. 

However there is variation in the rate of convergence with moment order, $n_\mathrm{mom}^{\mathrm{max}}$, over the different reference states. If we consider a system to be converged when the difference between a given moment order and the large moment limit is less than $10$meV, we find HF converges by the 5th moment, while DFT functionals converge between the 11th and 15th moment (depending on the fraction of exact exchange). This rapid convergence with HF references is largely echoed over the full $GW$100 test set, with another example, carbon monoxide (CO), shown explicitly in Appendix~\ref{app:momcon}. This fast convergence for these small molecules reflects that the HF initial IP is already a decent approximation, with an MAE of only $\sim$0.72eV between converged $G_0W_0$@HF and the mean-field HF values from Koopmans theorem. This therefore requires a lower dynamical resolution of the self-energy (and hence maximum moment order) in order to resolve the changes from this initial reference due to the correlations. 
We also note that a HF reference was previously also found to provide more accurate gaps and frontier excitations than DFT references over this test set~\cite{Bintrim2021}. RPA screening is not found to provide much of a change to the calculated IPs compared to TDA, as long as some degree of (first-order) exact exchange is included in the reference state. These observations may however not extend to more extended or solid-state systems, where the polarizability is larger and DFT is likely to be a better starting point. Nevertheless, for this reason we exclusively consider a HF reference over the remaining $GW$100 test set in subsequent sections of this work.

In Fig.~\ref{fig:borane_gw} we can now consider the moment convergence for the self-consistent adaptations of the moment conserving $GW$ implementation for this exemplar molecule. We explicitly show the reference independence of both the fs$GW$ variant, and the fully self-consistent sc$GW$, with both HF and PBE references giving identical results at each order. This sc$GW$ value is also shown to converge to good agreement with two reference values from the literature~\cite{Caruso2016,Wen2024}, with remaining discrepancy likely to result from the analytic continuation of these reference values. Single shot and evGW calculations are observed to converge significantly faster than the self consistent implementations, with the ev$GW$ only providing a small shift compared to $G_0W_0$. Once again, the results and convergence rates are also observed to remain relatively consistent between RPA and TDA screening.}


\subsection{Computational details}


While the original $GW$100 benchmarked $G_0W_0$ on a PBE reference state, here we exclusively consider a Hartree--Fock (HF) initial reference state, as discussed in the previous section. 
We use a def2-TZVPP basis set and matching default density fitting basis,
along with the def2-TZVPP effective core potential for the atoms
Rb, Ag, I, Cs, Au, and Xe, with the \texttt{PySCF} codebase used to set up these systems~\cite{PySCF2017, PySCF2020}. \toadd{We ensure that all molecules and variants are performed up to a moment order of 9, ensuring that the most costly fully self-consistent variants can be completed for the largest systems. While we expect the $G_0W_0$ and ev$GW$ results to be converged with this, the fs$GW$ and sc$GW$ results may have a small amount of residual moment error, which we would expect to result in a small overestimation of the resulting IP.}
Across the different self-consistencies, the convergence threshold was set to \toadd{$5$ meV} for the energies and $10^{-4}$ for the maximum absolute difference in any element of the moments over all orders. Along with these thresholds, a constant DIIS space of $12$ was employed,
while all RPA calculations used $32$ points in the numerical integration of Eq.~\ref{eq:Gmain}. Apart from the sc$GW$ calculations, the density fitting basis was compressed to a natural auxiliary basis (see Sec.~\ref{sec:Wmom}), with $\epsilon_{\textrm{NAF}}=10^{-5}$.

To benchmark the results of frontier excitations for our methods across the $GW$100 set, we also needed to establish a `ground truth' method which we can rely on for an accurate comparison.
The set initially relied on the difference of ground-state CCSD(T) results from an unrestricted Hartree Fock (UHF) reference between the neutral and ionized systems, seen to be a reliable, highly accurate method. Since the initial work, CCSD(T) with RHF along with EOM-CCSD calculations have been used for comparison. For this work, the CCSD(T) results for the $GW$100 set were recalculated using {\tt ORCA} with a frozen core approximation and a UHF reference~\cite{Neese2020}. These results are in good agreement with the original CCSD(T)~\cite{VanSetten2015, Krause2015}, EOM-CCSD of Lange et. al.~\cite{Lange2018} and the CCSD(T) results of Bruneval et al.~\cite{Bruneval2021}, with a mean absolute error in between our CCSD(T) results and those of Bruneval et al. across the set of only 0.04eV. 
However when comparing across these $GW$ and other results \cite{Schambeck2024, Bruneval2021} and different coupled-cluster based benchmarks, a common outlier appears, MgO. This has an error in our CCSD(T) reference IP of 1.8eV to the EOM-CCSD and 0.4eV to the Bruneval et al. CCSD(T) results. This system has been highlighted before as a particular challenge for coupled-cluster~\cite{Bauschlicher2018}, and we have strong belief that the benchmark excitation energy is untrustworthy for this system, and thus is excluded from the dataset when comparing results to the CCSD(T) reference data.

\subsection{$GW$100 results} \label{sec:resultsgw100}

\ifjcp
\begin{figure*}[ht]
    \centering
    \includegraphics[width=\textwidth]{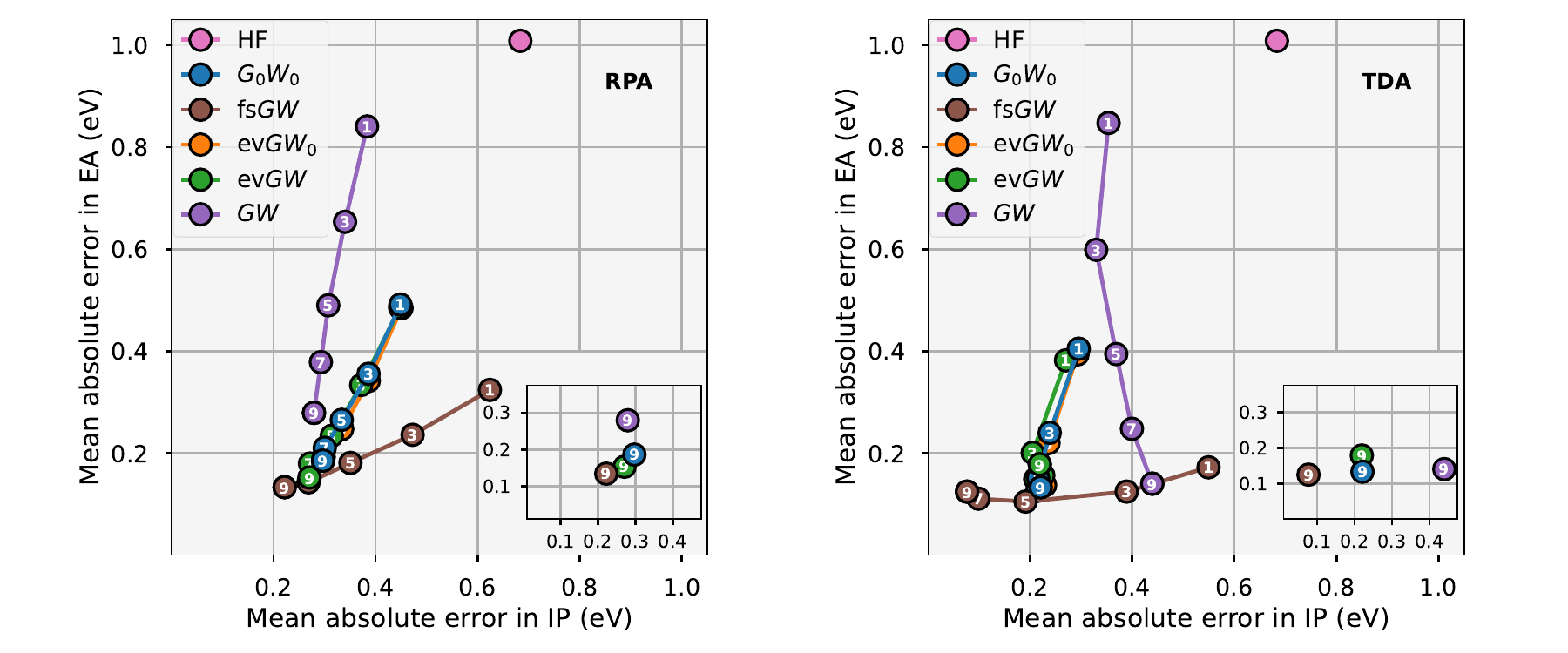}
    \caption{
        Convergence of the \ac{mae} for the IP ($x$-axis) and EA ($y$-axis) across the $GW$100 test set with respect to CCSD(T) reference values for many of the self-consistent $GW$ approaches considered. Numbers in the circles represent the maximum order of the conserved self-energy moments
        for each method, showing convergence to the full-frequency limit. The left plot shows the aggregated results for \ac{rpa} screening, while the right plot is for \ac{tda} screening. Inset of each plot depicts the aggregated MAE for moment order $9$ across the different moment-conserving GW variants.
    } 
    \label{fig:gw100_convergence}
\end{figure*}
\else
\begin{figure}[ht]
    \centering
    \includegraphics[width=\textwidth]{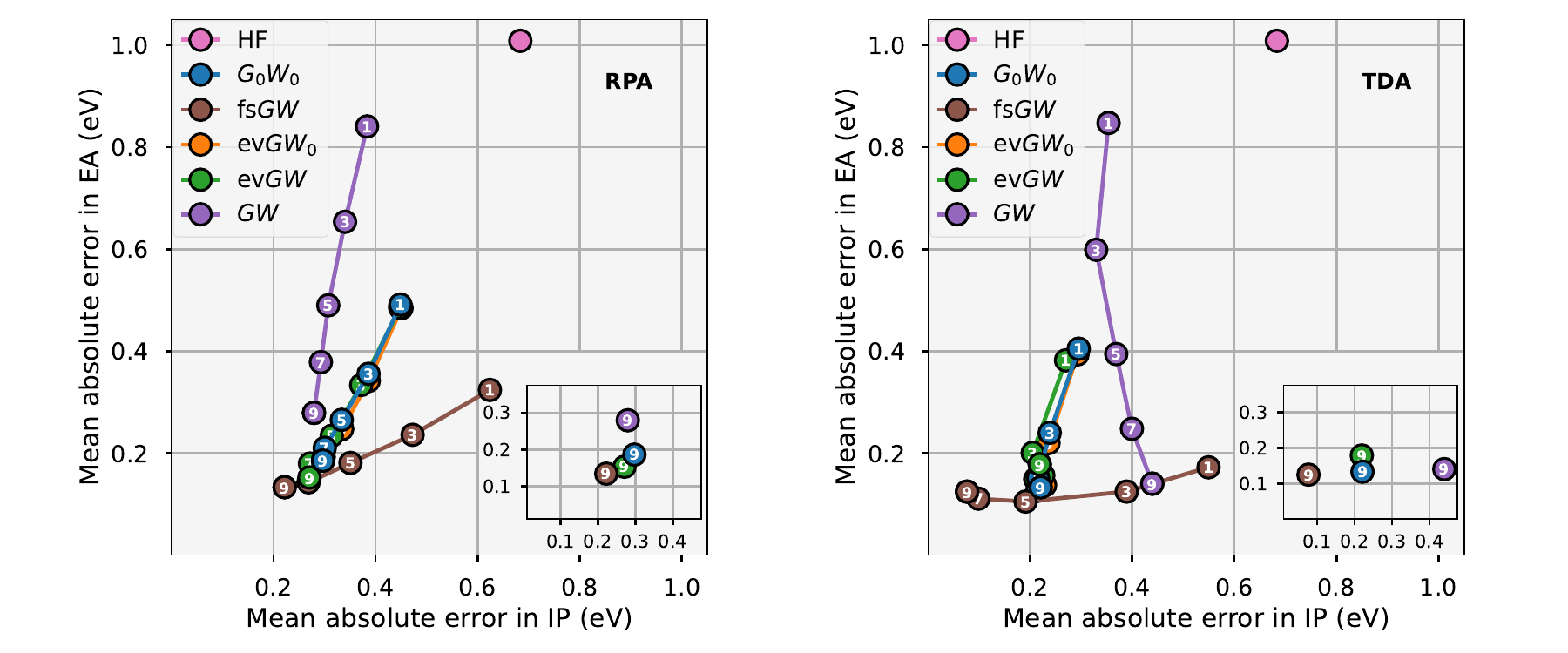}
    \caption{
        Convergence of the \ac{mae} for the IP ($x$-axis) and EA ($y$-axis) across the $GW$100 test set with respect to CCSD(T) reference values for many of the self-consistent $GW$ approaches considered. Numbers in the circles represent the maximum order of the conserved self-energy moments
        for each method, showing convergence to the full-frequency limit. The left plot shows the aggregated results for \ac{rpa} screening, while the right plot is for \ac{tda} screening. Inset of each plot depicts the aggregated MAE for moment order $7$ across the different moment-conserving GW variants.
    } 
    \label{fig:gw100_convergence}
\end{figure}
\fi

\cref{fig:gw100_convergence} shows the \ac{mae} in the \ac{ip} and \ac{ea}
aggregated over the whole $GW$100 test set for RPA screening (left) and TDA screening (right),
compared to explicit electron addition and removal calculations at the CCSD(T) level. Since these benchmarks are taken with respect to a CCSD(T) benchmark,
they are not expected to converge to zero error in the limit of high moment order, but rather the intrinsic error of that particular method.
In each plot, the $x$-axis indicates the \ac{mae} for the \ac{ip},
and the $y$-axis for the \ac{ea}, with the progression toward the origin indicating a convergence over the aggregated data to the CCSD(T) level.
While all methods display good convergence with increasing moment order to the inherent accuracy of the method,
$GW_0$ results for $n_{\mathrm{mom}}^{\mathrm{max}}=1$ are noteworthy as an outlier for being extremely poor for both \ac{tda} and \ac{rpa} screening. When examined on a per system basis, there are 4-8 systems that are significant outliers for this method, with the \ac{ip} outliers forming a subset of the \ac{ea} outliers. These systems are no longer outliers in higher moment orders, with relatively small distributions over the set as $n_{\mathrm{mom}}^{\mathrm{max}}$ increases, with the moment approximation converging to close to an exact full-frequency treatment of the method. We find that by a moment order of approximately seven, the moment approximation error is largely converged, and far smaller than the intrinsic error of the method, resulting in diminishing returns to further increase the moment order. The change in these frontier excitation energies between moment order of seven and nine is almost always significantly less than 100 meV, which is near the scatter of traditional $GW$ implementations due to grid convergence or differences in the convolution treatment~\cite{VanSetten2015}.

When comparing results between the \ac{tda} and \ac{rpa} screening, we find that particularly for the IP, \ac{tda} screening is seen to perform better than \ac{rpa} screening for all $GW$ variants, despite containing significantly fewer diagrammatic contributions. This has also been noted for molecular systems previously in Ref.~\onlinecite{Bintrim2021}. The significantly simpler \ac{tda} leads to a much more efficient (and parallelizable) algorithm,
resulting in a very viable method for charged excitations in molecules. \toadd{This good performance of the \ac{tda} screening must result from favorable error cancellation. This is likely to arise from the reduction in the IP from TDA screening treatment compensating for the overestimation arising from the Hartree--Fock reference state. It is likely that such a favorable cancellation may not appear in alternate reference states, where e.g. $G_0W_0$@PBE calculations have been found to underestimate the IP with RPA screening over this test set~\cite{Caruso2016}, and the \ac{tda} may compound rather than alleviate this error. Similarly, the importance of plasmonic-like collective excitations in more polarizable extended systems is likely to also necessitate the enhanced diagrammatic content of RPA screening in GW.}

\ifjcp
\begin{figure*}
    \centering
    \includegraphics[width=\textwidth]{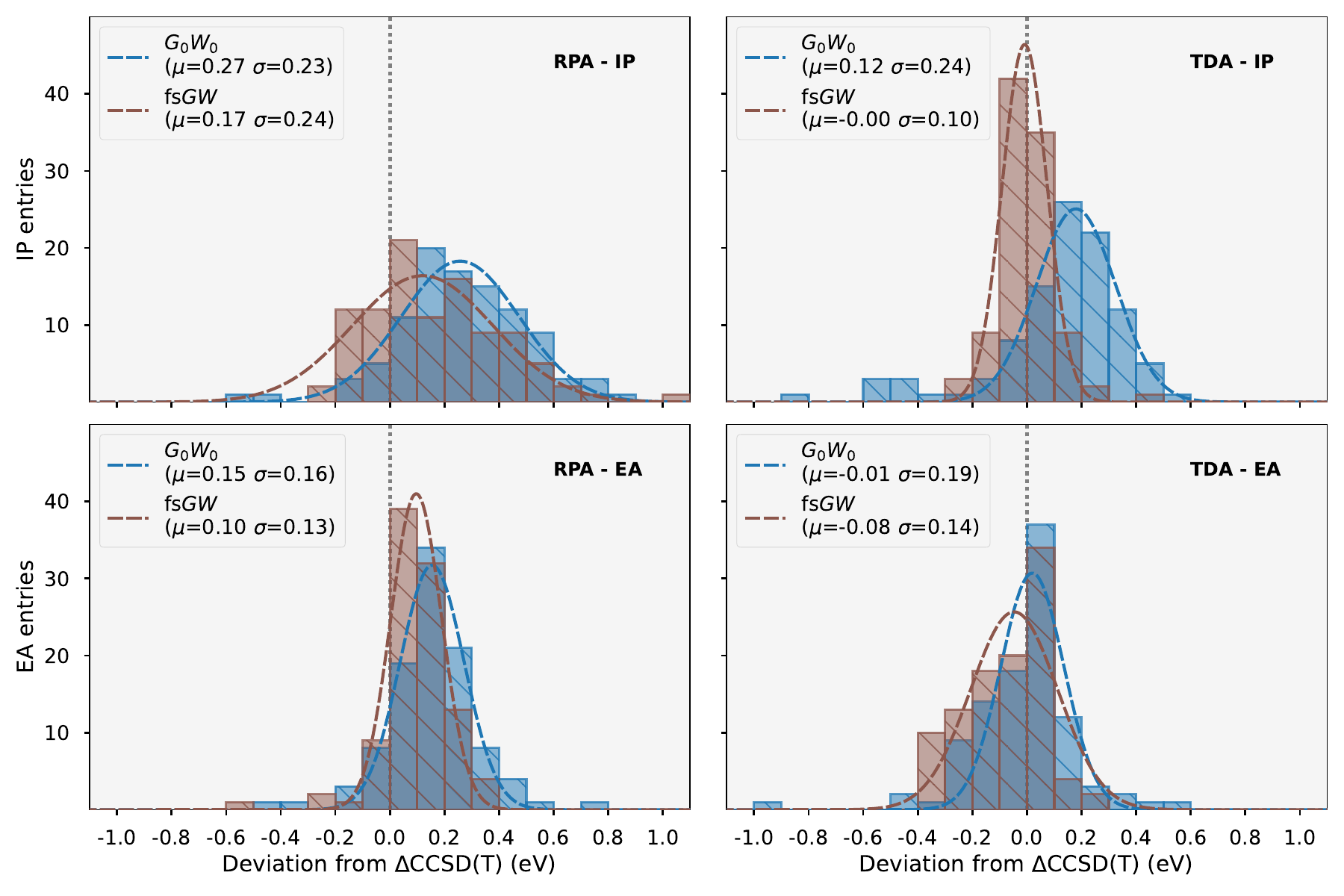}
    \caption{Histograms of the signed errors for the IP and EA with $G_0W_0$ and fs$GW$ relative to CCSD(T) over the GW100 set. Results presented for maximum moment order of 9 with Gaussians fit to the error distribution shown as dashed lines. The left plot shows RPA screening, right shows TDA screening with \ac{ip} results shown top and \ac{ea} bottom. The mean signed error $(\mu)$ and the standard deviation $(\sigma)$ for each method are shown in the legend.}
    \label{fig:gw100_hist}
\end{figure*}
\else
\begin{figure}
    \centering
    \includegraphics[width=\textwidth]{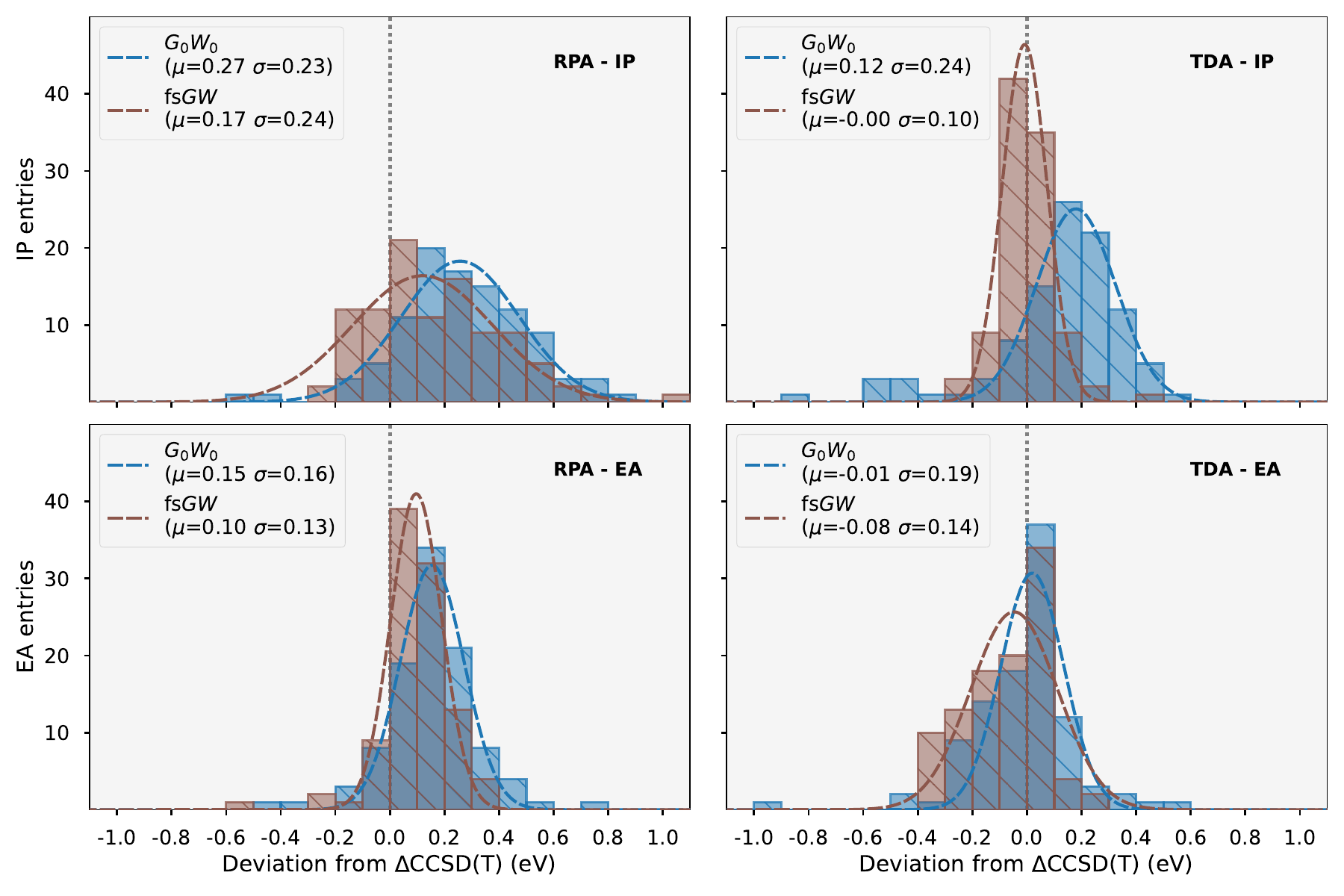}
    \caption{Histograms of the signed errors for the IP and EA with $G_0W_0$ and fs$GW$ relative to CCSD(T) over the GW100 set. Results presented for maximum moment order of 9 with Gaussians fit to the error distribution shown as dashed lines. The left plot shows RPA screening, right shows TDA screening with \ac{ip} results shown top and \ac{ea} bottom. The mean signed error $(\mu)$ and the standard deviation $(\sigma)$ for each method are shown in the legend.}
    \label{fig:gw100_hist}
\end{figure}
\fi

We can also compare the self-consistent variants over this test set, we find that $G_0W_0$, ev$GW_0$ and ev$GW$ perform very similarly, indicating that the eigenvalue self-consistency seems to be having little effect in these systems, presumably due to its inability to change the orbitals from the reference HF state\toadd{, and with the HF reference already providing a reasonable approximation for these large gapped molecular systems, as discussed in Sec.~\ref{sec:example}}. Comparatively $GW_0$ and $GW$ approach convergence with respect to moment order differently, with an initially larger \ac{ip} than \ac{ea} \ac{mae}, however it converges to similar values to $G_0W_0$ and ev$GW$ variants for higher moments. This pathway is similarly seen within fs$GW$, however it rapidly converges to a lower \ac{mae} for both \ac{ip} and \ac{ea} with respect to moment order. No results are shown for qs$GW$ due to convergence issues associated with a small number of molecules at higher moment order, noting that unreliable convergence of qs$GW$ has been observed in the literature~\cite{Monino2023}.

To better consider the best performing self-consistent fs$GW$ variant, \cref{fig:gw100_hist} compares fs$GW$ and $G_0W_0$ for moment order 9, again benchmarked against CCSD(T), as histograms of the errors for \ac{ip} (top) and \ac{ea} (bottom) over all molecules in the set. For TDA screening of the first \ac{ip}, fs$GW$ is seen to perform significantly better than $G_0W_0$ with an average difference of only $-0.4$meV and a narrow spread in results given by a standard deviation of $0.1$eV. Only one system, TiF4, deviates beyond $0.3$eV from the CCSD(T) result, noting that this system also appears as an outlier for a similar method described in Ref.~\onlinecite{Bruneval2021}, casting some doubt on the reliability of the benchmark CCSD(T) for that molecule. The \ac{ea} results are more similar between the two methods with small mean errors and standard deviation. For RPA screening, both methods are seen to predominately overestimate the energies, but both the \ac{ea} and \ac{ip} fs$GW$ results provide a small but significant improvement in the mean error over $G_0W_0$. \toadd{While these fs$GW$ results are encouraging, further benchmarking of the method is required over a larger set of systems and materials, noting that the different rate of convergence with respect to moment order also requires care for a faithful comparison.}

\begin{table}[H]
\caption{Mean absolute error (\ac{mae}), mean signed error (MSE) and standard deviation (STD) for $G_0W_0$@$HF$, ev$GW$, fs$GW$ and sc$GW$ across the GW100 test set. All results shown in meV at $n_{\mathrm{mom}}^{\mathrm{max}}=9$.}
\begin{center}

\begin{tabular}{|p{0.14\linewidth}?P{0.12\linewidth}P{0.12\linewidth}P{0.12\linewidth}?P{0.12\linewidth}P{0.12\linewidth}P{0.12\linewidth}|}
\multicolumn{1}{l}{}  & \multicolumn{3}{c?}{\large\textrm{IP}} &  \multicolumn{3}{c}{\large\textrm{EA}} \\ \hline 

\multicolumn{1}{|c?}{\textbf{RPA}} & \multicolumn{1}{c|}{MAE} & \multicolumn{1}{c|}{MSE} & \multicolumn{1}{c?}{STD} & \multicolumn{1}{c|}{MAE} & \multicolumn{1}{c|}{MSE} & \multicolumn{1}{c|}{STD} \\
\hline
$G_0W_0$ & \scriptsize 297 & \scriptsize 266 & \scriptsize 233 & \scriptsize 186 & \scriptsize 153 & \scriptsize 163 \\
fs$GW$ & \scriptsize 222 & \scriptsize 168 & \scriptsize 238 & \scriptsize 134 & \scriptsize 99 & \scriptsize 131 \\
ev$GW$ & \scriptsize 271 & \scriptsize 235 & \scriptsize 215 & \scriptsize 153 & \scriptsize 153 & \scriptsize 169 \\
sc$GW$ & \scriptsize 279 & \scriptsize -157 & \scriptsize 301 & \scriptsize 279 & \scriptsize 271 & \scriptsize 166 \\ \hline 

\multicolumn{1}{|c?}{\textbf{TDA}} & \multicolumn{1}{c}{ } & \multicolumn{1}{c}{ } & \multicolumn{1}{c?}{ } & \multicolumn{1}{c}{ } & \multicolumn{1}{c}{ } & \multicolumn{1}{c|}{ } \\
\hline
$G_0W_0$ & \scriptsize 221 & \scriptsize 119 & \scriptsize 239 & \scriptsize 134 & \scriptsize -12 & \scriptsize 190 \\
fs$GW$ & \scriptsize 77 & \scriptsize 0 & \scriptsize 103 & \scriptsize 125 & \scriptsize -83 & \scriptsize 142 \\
ev$GW$ & \scriptsize 219 & \scriptsize 38 & \scriptsize 264 & \scriptsize 179 & \scriptsize -108 & \scriptsize 213 \\
sc$GW$ & \scriptsize 439 & \scriptsize -439 & \scriptsize 227 & \scriptsize 141 & \scriptsize 112 & \scriptsize 142 \\ \hline 
\end{tabular}
\end{center}
 \label{tab:gw100_mae_mse_std}
\end{table}

Finally, Tab.~\ref{tab:gw100_mae_mse_std} summarizes key statistics for the performance of the main self-consistent moment-conserving $GW$ methods across the $GW$100 test set for maximum moment order \toadd{nine}. These aggregated statistics reinforce the trend of TDA screening performing better than RPA over these molecular systems, with fs$GW$@TDA the best approach for both IP and EA estimates. Overall, we find that the trend of $GW$ methods is to overestimate both IP and EA values in general, such that the errors partially cancel for fundamental gaps. While this could be attributed to the HF starting point in this work, we find this persisting even for self-consistent variants. This results in the errors in the fundamental gap in general being smaller than the errors in the individual IP and EA~\cite{Bruneval2021}. 
Finally, we note that neither fully self-consistent $GW$ (despite significant additional cost) nor ev$GW$ provide clear tangible benefit over $G_0W_0$@HF when aggregated over these systems, further highlighting the value in alternate (albeit heuristic) self-consistent formulations such as fs$GW$. These values are also similar to those found in related studies over the IPs in Ref. ~\onlinecite{Wen2024}. However, in their work, the fully self-consistent $GW$ implementation performs slightly more favorably than the single shot, \toadd{with a MAE for $G_0W_0$@HF and sc$GW$ of 0.35eV and 0.29eV respectively. In this work, the improvement is not as clear when compared with the same reference CCSD(T) results. This remaining small discrepancy may be due to residual moment order incompleteness, as seen for $n_{\mathrm{mom}}=9$ in Fig. ~\ref{fig:borane_gw}, which tends to result in an overestimation of the IPs relative to the infinite moment limit. In addition to this, differences in the temperature, numerical thresholds and implementations of analytic continuation can all be factors, as well as the fact that the previous study considered 92 of the 101 systems considered here. Nevertheless, the agreement is broadly good.} 

\section{Chlorophyll A molecular chromophore} \label{sec:chlorophyll}

The Chlorophyll A molecule is central to photosynthesis, and features a light-harvesting complex connected to an extended conjugated $\pi$-system in a chlorin macrocycle ring with a magnesium ion at its center. This chlorin ring bears similarities to porphyrin rings, with its delocalized electronic structure enabling strong absorption in the visible wave length, whose initial excitation creates excitons as a first step in photosynthesis. The states of the magnesium ion play a key role in modulating the electronic states and tuning the absorption spectrum.

A particularly appealing aspect of the moment-conserving approach to $GW$ is that even when the dynamics of the self-energy are truncated to a particular moment order, the final correlated Green's function is always obtained over all frequencies, without requiring excitations to be converged sequentially, targeting of a specified energy window, or to rely on analytic continuation techniques. Indeed, since the self-energy moments are integrals over all frequencies, there is no reason to necessarily expect the accuracy of the frontier excitations to not be maintained throughout the spectrum (or indeed improved if core excitations are more rapidly converged with moment order of the self-energy). This allows for the full spectrum to be obtained straightforwardly from Eq.~\ref{eq:spectrum}. 
We apply our moment-conserving $G_{0}W_{0}$ algorithm and compare \ac{rpa} to \ac{tda} screening for the chlorophyll molecular chromophore within an aug-cc-pVDZ basis set.
This system has 137 atoms with 2147 orbitals and 482 electrons,
all of which are correlated, with the structure being obtained from the RCSB Protein Database (PDB). 

\ifjcp
\begin{figure*}[ht]
    \centering
    \includegraphics[width=\textwidth]{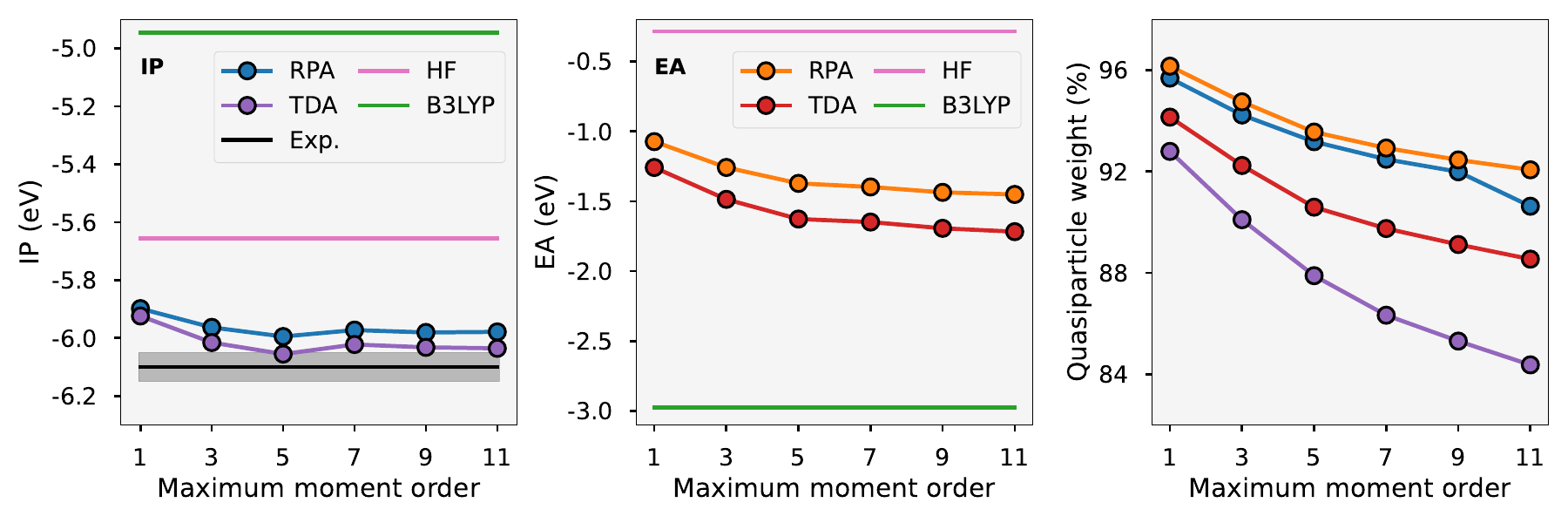}
    \caption{
        Convergence of the moment-conserving $G_0W_0$ \ac{ip} (left),
        \ac{ea} (middle),
        and their quasiparticle weights (right) with increasing conserved
        moment order for the Chlorophyll A molecule in an aug-cc-pVDZ basis set. Also included are the mean-field orbital energies corresponding to the IP and EA at the level of HF and DFT with a B3LYP functional. In addition, we include an experimental estimate of the IP energy from Ref.~\onlinecite{Shafizadeh2011}, with associated experimental uncertainty shown by the gray shaded region.}
    \label{fig:chlorophyll}
\end{figure*}
\else
\begin{figure}[H]
    \centering
    \includegraphics[width=\textwidth]{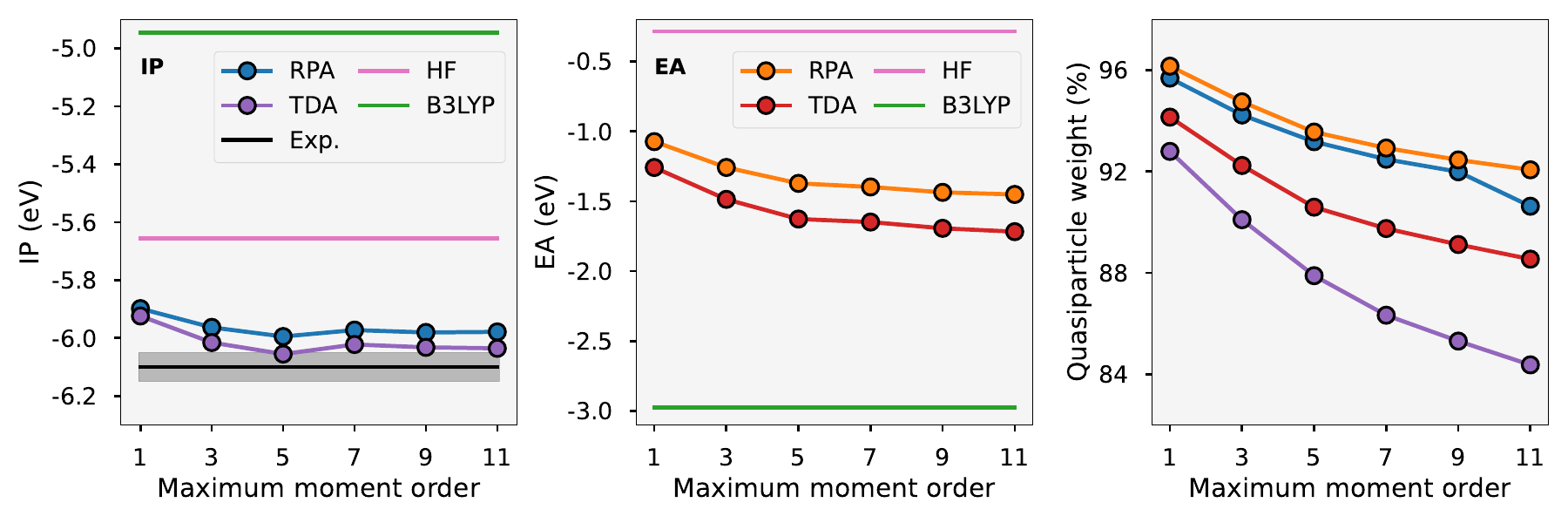}
    \caption{
        Convergence of the moment-conserving $G_0W_0$ \ac{ip} (left),
        \ac{ea} (middle),
        and their quasiparticle weights (right) with increasing conserved
        moment order for the Chlorophyll A molecule in an aug-cc-pVDZ basis set. Also included are the mean-field orbital energies corresponding to the IP and EA at the level of HF and DFT with a B3LYP functional. In addition, we include an experimental estimate of the IP energy from Ref.~\cite{Shafizadeh2011}, with associated experimental uncertainty shown by the gray shaded region.}
    \label{fig:chlorophyll}
\end{figure}
\fi

In Fig.~\ref{fig:chlorophyll} we show the convergence of the \ac{ip}, \ac{ea}, and their respective quasiparticle weights (i.e. residue of the pole in the spectrum) with respect to maximum conserved self-energy moment order. This demonstrates a strong convergence of the IP and EA positions, but a slower convergence of the quasiparticle weight. As the maximum moment order is increased, the total number of excitations grows linearly, and the proliferation of low-weighted excitations likely contributes to this slower convergence compared to the excitation energies. Perhaps surprisingly, we find that the TDA screened frontier excitations have a lower quasi-particle weight compared to the RPA screening, indicating a larger effect of correlation-induced splitting of these excitations compared to RPA, despite a smaller diagrammatic content of the screening physics. Overall, there is indeed some relatively substantial renormalization of these frontier excitations due to the correlation, with quasi-particle weights adjusted from a single particle picture by a factor of $\sim0.9$.

Returning to the IP energy, we can compare to an experimental result of $6.1\pm 0.05$eV from a pump-probe evaporation experiment~\cite{Shafizadeh2011}. We find the $G_0W_0$ results within $0.04$eV (TDA) and $0.11$eV (RPA) of this value, with the TDA inside the experimental uncertainty. However, both of these discrepancies compared to experiment are also likely within the error associated with the remaining basis set incompleteness error of the $G_0W_0$ predictions~\cite{Loos2020b}. Leaving this aside, we again note that the TDA screening is giving marginally better estimates of the IP compared to experiment, despite the significantly larger and more polarizable system that chlorophyll represents compared to the relatively small molecular systems of the $GW$100 test set, which would be expected to support plasmonic-like low-energy excitations only present in RPA screening. We find that both TDA and RPA screening is substantially more accurate compared to the HF and DFT@B3LYP estimates of the IP, with the B3LYP IP in error by 1.2eV compared to experiment, while for the EA the discrepancy to the $G_0W_0$ results are even larger. 

\ifjcp
\begin{figure*}[ht]
    \centering
    \includegraphics[width=0.49\textwidth,trim={30cm 0cm 0cm 40cm},clip]{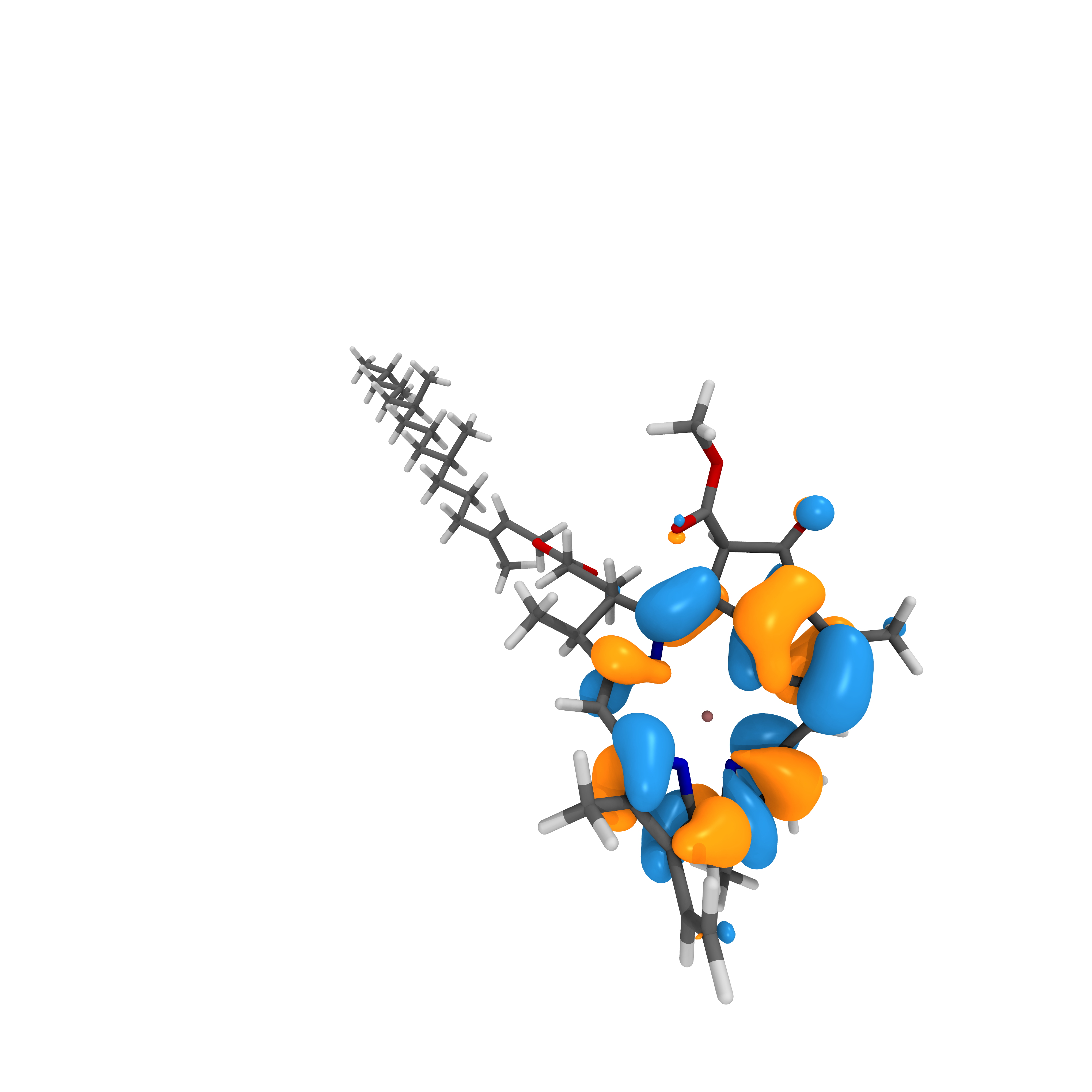}
    \includegraphics[width=0.49\textwidth,trim={30cm 0cm 0cm 40cm},clip]{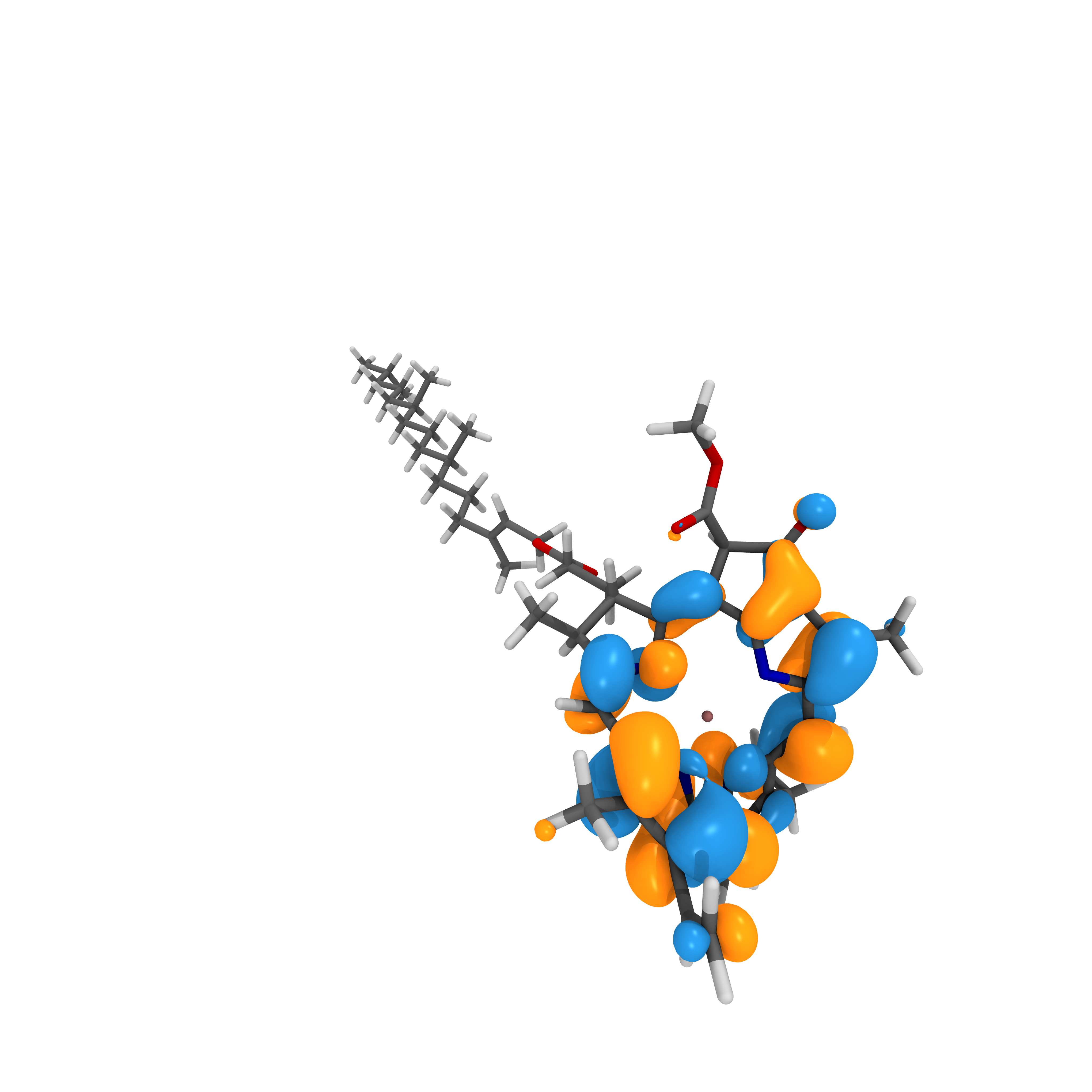}
    \caption{
        Dyson orbitals for the HOMO (left) and LUMO (right) for
        chlorophyll in an aug-cc-pVDZ basis set
        for the moment-conserving $G_0W_0$ method,
        using \ac{rpa} and \ac{tda} screening,
        with maximum moment order $n_{\mathrm{mom}}^{\mathrm{max}}=11$.
    }
    \label{fig:chlorophyll_orbitals}
\end{figure*}
\else
\begin{figure}[ht]
    \centering
    \includegraphics[width=0.49\textwidth,trim={30cm 0cm 0cm 40cm},clip]{homo.png}
    \includegraphics[width=0.49\textwidth,trim={30cm 0cm 0cm 40cm},clip]{lumo.png}
    \caption{
        Dyson orbitals for the HOMO (left) and LUMO (right) for
        chlorophyll in an aug-cc-pVDZ basis set
        for the moment-conserving $G_0W_0$ method,
        using \ac{rpa} and \ac{tda} screening,
        with maximum moment order $n_{\mathrm{mom}}^{\mathrm{max}}=11$.
    }
    \label{fig:chlorophyll_orbitals}
    \end{figure}
\fi

The eigenvectors of $\bm{\tilde{H}}$ corresponding to the IP and EA in these approaches also provide information on the spatial character of these HOMO and LUMO excitations for the chlorophyll system. These Dyson orbitals are shown in \cref{fig:chlorophyll_orbitals}, according the the moment-conserving $GW$ calculations with maximum moment order $n_{\mathrm{mom}}^{\mathrm{max}}=11$, and dictate the character of the fundamental absorption processes in the initial excitation, where charge transfer from the HOMO to the LUMO is expected. While both Dyson orbitals are localized on the $\pi$-conjugated system of the chlorin ring, the LUMO extends further over the nearby carbonyl group and peripheral substituents. This shows that the exciton is expected to be initially delocalized over the chlorin ring before transferring energy through the light-harvesting complex to the reaction center, as also found in other studies~\cite{Hashemi2021, Forster2022b}.

\ifjcp
\begin{figure*}
    \centering
    \includegraphics[width=\textwidth]{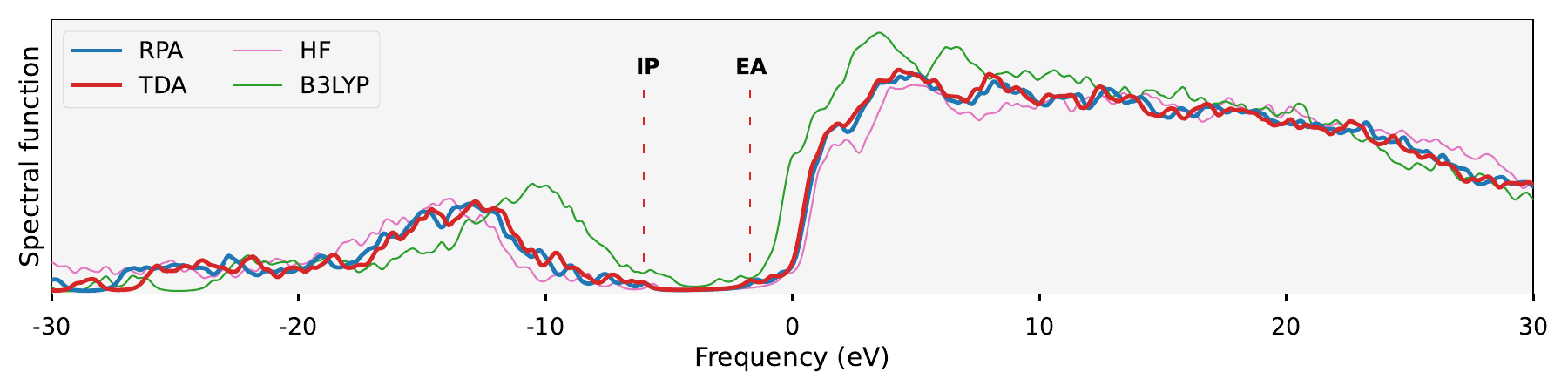}
    \caption{
        Spectral function for Chlorophyll A in an aug-cc-pVDZ basis set
        for the moment-conserving $GW$ method,
        using \ac{rpa} and \ac{tda} screening,
        with maximum moment order $n_{\mathrm{mom}}^{\mathrm{max}}=11$.
        Thinner lines show the Hartree--Fock and B3LYP spectral functions. IP and EA locations are shown for moment-conserving $G_0W_0$@TDA.}
    \label{fig:chlorophyll_spectra}
\end{figure*}
\else
\begin{figure}
    \centering
    \includegraphics[width=\textwidth]{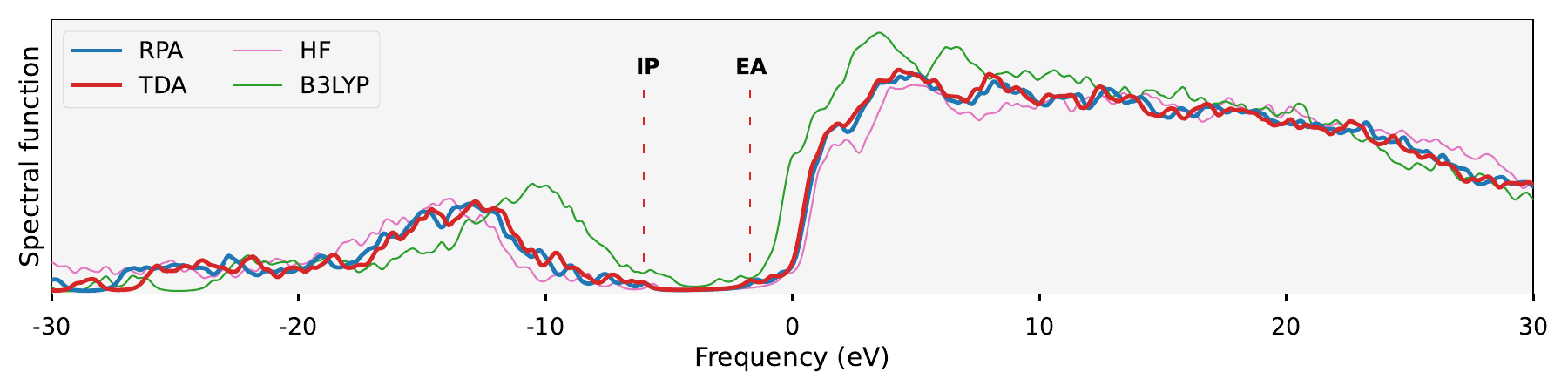}
    \caption{
        Spectral function for Chlorophyll A in an aug-cc-pVDZ basis set
        for the moment-conserving $GW$ method,
        using \ac{rpa} and \ac{tda} screening,
        with maximum moment order $n_{\mathrm{mom}}^{\mathrm{max}}=11$.
        Thinner lines show the Hartree--Fock and B3LYP spectral functions. IP and EA locations are shown for moment-conserving $G_0W_0$@TDA.}
    \label{fig:chlorophyll_spectra}
\end{figure}
\fi

In \cref{fig:chlorophyll_spectra} we also plot the spectrum over a much wider spectral range ($n_{\mathrm{mom}}^{\mathrm{max}}=11$), showing a qualitative similarity between the TDA and RPA screened spectrum with a broad band of excitations. The plot also includes the Hartree--Fock and B3LYP spectral functions, shown as fainter lines.
The fundamental gap of the spectrum is found to be $4.32$eV (TDA) and $4.53$eV (RPA) -- in the near UV rather than visible range -- demonstrating the importance of excitonic binding in computing the correct optical gap. Extensions of this moment-conserving approach towards the Bethe--Salpeter equation are underway in order to include these effects for optical processes. Additionally, the differences in quasiparticle weight seen in figure \ref{fig:chlorophyll} is less evident in the spectrum due to the broad band of higher weight excitations at larger energies. Nevertheless, the results for the charged spectrum are in good agreement with experiment, and demonstrate a scalable $GW$ approach with full spectral information. 

\section{Conclusions and outlook}

In this work we have extended the recently introduced moment-conserving framework for $GW$, demonstrating efficiency improvements, parallelism and extensions enabling a wide array of self-consistent adaptations to be reliably converged. The framework avoids analytic continuation and a number of other technical choices in most other $GW$ implementations concerning a choice of grids, temperatures, strategy for the convolution or quasiparticle approximations, relying instead solely on a convergence with the number of directly computed spectral moments in which the self-energy is expanded. These moments are then represented via a compact upfolded self-energy, enabling the full spectrum of excitations to be found in a single-shot diagonalization step.

Across the $GW$100 molecular test set we show that a diverse range of self-consistent formulations of $GW$ can be converged with respect to this moment approximation, reaching aggregated errors with respect to high level references in agreement with entirely separate grid-based $GW$ implementations~\cite{Wen2024, Caruso2016, Bruneval2021}. Comparing across self-consistent variants, we find that a self-consistency just at the level of the Fock matrix and chemical potential (which we dub `Fock self-consistent $GW$, fs$GW$) is particularly effective, while reducing the cost compared to a fully dynamical self-consistency. Furthermore, we corroborate previous indications that (somewhat counter-intuitively) TDA rather than RPA screening provides a higher level of accuracy in the position of these frontier excitations for these molecular systems. This TDA screening also significantly simplifies the complexity of the moment-based $GW$ calculations compared to RPA screening (noting the same $\mathcal{O}[N^4]$ formal scaling).
We also consider the more extended Chlorophyll A molecule, where again we find the first IP with TDA screening to be more accurate than RPA when compared to experimental values. However, both are very close to these experimental values and other uncertainties may need more scrutiny before conclusive statements can be made as to the most appropriate screening in this system. We demonstrate that we can find these IP and EA Dyson orbitals, as well as the full-frequency spectrum, which exhibits quantitative differences to DFT as expected.

Looking forward, we are extending the implementation to enforce $\mathbf{k}$-point symmetry for solid-state applications, which will allow for a broader investigation into the effectiveness of the fs$GW$ self-consistent scheme as a lower-cost approach compared to fully dynamical self-consistency. Furthermore, we will be able to investigate the limits of TDA screening in $GW$, noting that the expectation is that at some length scale the polarizability of the system will be sufficiently large to support higher-body RPA-like plasmonic charge fluctuations which will become essential for a faithful description. However, the `crossover' between these physical regimes and areas of applicability of TDA screening in both molecular and the solid state is still unclear.

We can also formulate the Bethe--Salpeter equation within this formally `static' moment-conserving picture, which will enable an efficient recasting of optical and excitonic phenomena in this framework. Finally, we note that the bottlenecks of the algorithm which prevents further scaling to larger systems is predominantly the memory bottleneck of the factorized Coulomb interaction ($\mathcal{O}[N^3]$), since our CPU scaling is only approximately a constant order of magnitude more than Hartree--Fock theory. Therefore, we will investigate the use of doubly factorized `tensor hypercontraction' techniques to remove this memory bottleneck, and enable a further step-change in the size of the systems which we can explore within these approaches.


\ifjcp\else
    \newpage
\fi

\section*{Code Availability}
The code for this project is fully open-source and available at \href{https://github.com/BoothGroup/momentGW}{https://github.com/BoothGroup/momentGW}, while the {\tt dyson} package for constructing upfolded moment-conserving Hamiltonians is available at \href{https://github.com/BoothGroup/dyson}{https://github.com/BoothGroup/dyson}.

\section*{Acknowledgements}
We sincerely thank Dominika Zgid for discussions on technical details of particle number conservation in sc$GW$. We thank Kemal Atalar for helpful comments on this manuscript. GHB acknowledges funding from the European Union's Horizon 2020 research and innovation programme under grant agreement No. 759063.
We are also grateful to the UK Materials and Molecular Modelling Hub for computational resources, which is partially funded by EPSRC (EP/P020194/1 and EP/T022213/1).


\ifjcp
%
\else
    \bibliographystyle{achemso}
\providecommand{\latin}[1]{#1}
\makeatletter
\providecommand{\doi}
  {\begingroup\let\do\@makeother\dospecials
  \catcode`\{=1 \catcode`\}=2 \doi@aux}
\providecommand{\doi@aux}[1]{\endgroup\texttt{#1}}
\makeatother
\providecommand*\mcitethebibliography{\thebibliography}
\csname @ifundefined\endcsname{endmcitethebibliography}
  {\let\endmcitethebibliography\endthebibliography}{}

\fi

\appendix
\setcounter{equation}{0}
\section{Schematic self-consistent moment-$GW$ algorithms}
\label{app:algos}
\renewcommand{\theequation}{A\arabic{equation}}

In this appendix, we provide pseudocode for the remaining self-consistent moment-conserving $GW$ algorithms discussed in the main text.

\begin{algorithm}[H]
    \begin{algorithmic}
        \Function{evGW}{
            $\bm{\upepsilon}$,
            $\mathbf{C}$,
            $\mathbf{V}^{\mathrm{AO}}$,
            $n_{\mathrm{mom}}^{\mathrm{max}}$,
            \texttt{g0},
            \texttt{w0}
        }
            \State $f_{p,p}, \Sigma_{\infty,pq} \gets \epsilon_{p} \delta_{pq}$ \Comment{\cref{eq:StaticSE}}
            \State $V_{Q, ia} \gets \sum_{\alpha \beta} V^{\mathrm{AO}}_{Q, \alpha \beta} C_{\alpha i}^{*} C_{\beta a}$ \Comment{Distribute over $(i,a)$}
            \State $\hat{V}_{Q, px} \gets \sum_{\alpha \beta}  V^{\mathrm{AO}}_{Q, \alpha \beta} C_{\alpha p}^* C_{\beta x}$ \Comment{Distribute over $x$}
            \State $\epsilon_{p}^{G} \gets \epsilon_{p}$
            \State $\epsilon_{p}^{W} \gets \epsilon_{p}$
            \While{\textbf{not} converged}
                \State $\mathbf{D} \gets \bm{\upepsilon}^{W}$
                \Comment{\cref{eq:energy_differences}}
                \State $\tilde{\bm{\upeta}}^{(n)} \gets \mathbf{V}, \mathbf{D}$
                \Comment{\cref{sec:rpa} or \cref{sec:tda}}
                \State $\mathbf{W}^{(n)} \gets \mathbf{V}, \hat{\mathbf{V}}, \tilde{\bm{\upeta}}^{(n)}$
                \Comment{\cref{eq:w_moments}}
                \State $\bm{\Sigma}^{(n, <)}, \bm{\Sigma}^{(n, >)}  \gets \bm{\upepsilon}^{G}, \mathbf{W}^{(n)}$
                \Comment{\cref{eq:convolution_occ} and \cref{eq:convolution_vir}}
                \State $\mathbf{\tilde{H}} \gets f_{p,p}, \bm{\Sigma}_\infty, \bm{\Sigma}^{(n, <)}, \bm{\Sigma}^{(n, >)}$
                \Comment{\texttt{dyson.MBLSE}}
                \State $\bm{\chi} \mathbf{E} \bm{\chi}^\dagger \gets \mathbf{\tilde{H}}$
                \If{\textbf{not} \texttt{g0}}
                    \State $\bm{\upepsilon}^{G} \gets \mathbf{E}$
                    \Comment{\cref{eq:quasiparticle_energy_update}}
                \EndIf
                \If{\textbf{not} \texttt{w0}}
                    \State $\bm{\upepsilon}^{W} \gets \mathbf{E}$
                    \Comment{\cref{eq:quasiparticle_energy_update}}
                \EndIf
            \EndWhile
            \State \Return{}
        \EndFunction
    \end{algorithmic}
    \caption{Moment based \Acf{evgw} implementation. \texttt{g0} and \texttt{w0} flag whether the Green's function and/or screened Coulomb interaction are excluded from self-consistency.}
    \label{alg:evgw}
\end{algorithm}

\begin{algorithm}[H]
    \begin{algorithmic}
        \Function{qsGW}{
            $\bm{\upepsilon}$,
            $\mathbf{C}$,
            $\mathbf{V}^{\mathrm{AO}}$,
            $n_{\mathrm{mom}}^{\mathrm{max}}$
        }
            \State $F_{pq} \gets \epsilon_{p} \delta_{pq}$ \Comment{\cref{eq:StaticSE}}
            \State $\bm{\tilde{H}} \gets$ \Call{$G_0W_0$}{
                $\bm{\upepsilon}$,
                $\mathbf{C}$,
                $\mathbf{V}^{\mathrm{AO}}$,
                $n_{\mathrm{mom}}^{\mathrm{max}}$
            }
            \Comment{\cref{alg:g0w0}}
            \State $\bm{\upvarepsilon}, \mathbf{v} \gets \bm{\tilde{H}}$ 
            \State $\epsilon_{p}^{\mathrm{QP}} \gets \epsilon_{p}$
            \While{\textbf{not} converged}
                \State $\mathbf{V}^{\Sigma} \gets \bm{\upvarepsilon}, \mathbf{v}$
                \Comment{\cref{eq:qsgw_se_eta} or \cref{eq:qsgw_se_srg}}
                \State $\bm{\upepsilon}^{\mathrm{QP}}, \mathbf{C} \gets$ \Call{SCF}{
                    $\mathbf{F}$,
                    $\mathbf{V}^{\Sigma}$
                }
                \Comment{\cref{eq:qp_scf}}
                \State $\bm{\tilde{H}} \gets$ \Call{$G_0W_0$}{
                    $\bm{\upepsilon}^{\mathrm{QP}}$,
                    $\mathbf{C}$,
                    $\mathbf{V}^{\mathrm{AO}}$,
                    $n_{\mathrm{mom}}^{\mathrm{max}}$
                }
                \State $\bm{\upvarepsilon}, \mathbf{v} \gets \bm{\tilde{H}}$ 
            \EndWhile
            \State \Return{}
        \EndFunction
    \end{algorithmic}
    \caption{Moment-based \Acf{qsgw} implementation}
    \label{alg:qsgw}
\end{algorithm}

\begin{algorithm}[H]
    \begin{algorithmic}
        \Function{scGW}{
            $\bm{\upepsilon}$,
            $\mathbf{C}$,
            $\mathbf{V}^{\mathrm{AO}}$,
            $n_{\mathrm{mom}}^{\mathrm{max}}$,
            \texttt{g0},
            \texttt{w0}
        }
            \State $f_{p,p}, \Sigma_{\infty,pq} \gets \epsilon_{p} \delta_{pq}$ \Comment{\cref{eq:StaticSE}}
            \State $V_{Q, ia} \gets \sum_{\alpha \beta} V^{\mathrm{AO}}_{Q, \alpha \beta} C_{\alpha i}^{*} C_{\beta a}$ \Comment{Distribute over $(i,a)$}
            \State $\hat{V}_{Q, px} \gets \sum_{\alpha \beta}  V^{\mathrm{AO}}_{Q, \alpha \beta} C_{\alpha p}^* C_{\beta x}$ \Comment{Distribute over $x$}
            \State $\epsilon_{p}^{G} \gets \epsilon_{p}$
            \State $\epsilon_{p}^{W} \gets \epsilon_{p}$
            \State $\hat{\chi}_{\alpha\beta} \gets \delta_{\alpha\beta}$ 
            \While{\textbf{not} converged}
                \State $\mathbf{D} \gets \bm{\upepsilon}^{W}$
                \Comment{\cref{eq:energy_differences}}
                \State $\tilde{\bm{\upeta}}^{(n)} \gets \mathbf{V}, \mathbf{D}$
                \Comment{\cref{sec:rpa} or \cref{sec:tda}}
                \State $\mathbf{W}^{(n)} \gets \mathbf{V}, \hat{\mathbf{V}}, \tilde{\bm{\upeta}}^{(n)}$
                \Comment{\cref{eq:w_moments}}
                \State $\bm{\Sigma}^{(n, <)}, \bm{\Sigma}^{(n, >)} \gets \bm{\upepsilon}^{G}, \mathbf{W}^{(n)}$
                \Comment{\cref{eq:convolution_occ} and \cref{eq:convolution_vir}}
                \State $\mathbf{\tilde{H}} \gets f_{p,p}, \bm{\Sigma}_\infty, \bm{\Sigma}^{(n, <)}, \bm{\Sigma}^{(n, >)}$ \Comment{\texttt{dyson.MBLSE}}
                \State $\bm{\chi} \mathbf{E} \bm{\chi}^\dagger \gets \mathbf{\tilde{H}}$ 
                \State $\hat{\bm{\chi}} \gets \hat{\bm{\chi}}\bm{\chi}$ 
                \State $\bm{T} \gets \bm{C}\hat{\bm{\chi}}$ \Comment{Transformation AO $\rightarrow$ QMO }
                \If{\textbf{not} \texttt{g0}}
                    \State $\epsilon_{x}^{G} \gets E_{x}$
                    \State $\hat{V}_{Q, px} \gets \sum_{\alpha \beta}  V^{\mathrm{AO}}_{Q, \alpha \beta} C_{\alpha p}^* T_{\beta x}$ \Comment{Distribute $x$. }
                \EndIf
                \If{\textbf{not} \texttt{w0}}
                    \State $\epsilon_{x}^{W} \gets E_{x}$
                    \State $V_{Q, ia} \gets \sum_{\alpha \beta} V^{\mathrm{AO}}_{Q, \alpha \beta} T_{\alpha i}^{*} T_{\beta a}$ \Comment{Distribute $(i,a)$}
                \EndIf
            \EndWhile
            \State \Return{}
        \EndFunction
    \end{algorithmic}
    \caption{\Acf{scgw} implementation in \texttt{momentGW}.}
    \label{alg:scgw}
\end{algorithm}

\section{Carbon Monoxide moment convergence}
\label{app:momcon}

\toadd{The plots of Figs.~\ref{fig:CO_mf} and ~\ref{fig:CO_gw} show the moment order convergence of the results for $G_0W_0$ over different initial reference states, as well as for the various self-consistent variants, analogous to the results and discussion in Sec.~\ref{sec:example}.}

\ifjcp
\begin{figure*}
    \centering
    \includegraphics[width=\textwidth]{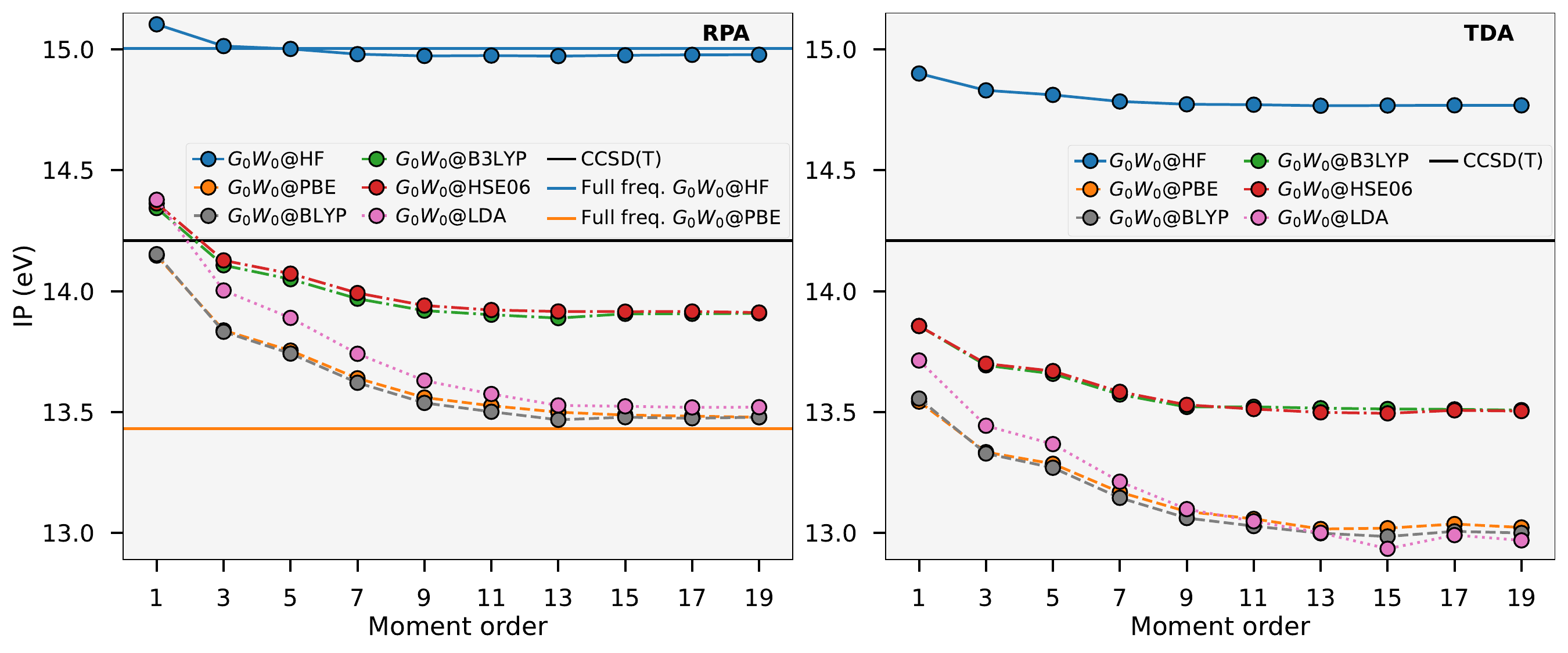}
    \caption{\toadd{Convergence of the IP of carbon monoxide (CO) with respect to the number of conserved moments ($n_\textrm{mom}^\textrm{max}$) of the self-energy in a def2-TZVPP basis set for single-shot $G_0W_0$, with RPA screening (left) and TDA (right). A range of mean-field starting points are considered, as well as reference values for RPA screening from {\tt PySCF}, implementing an $\mathcal{O}[N^6]$ full-frequency algorithm to remove any grid approximations \cite{doi:10.1021/acs.jctc.0c00704}. The remaining discrepancy likely comes from the diagonal approximation to the self-energy enforced in the reference values.}}
    \label{fig:CO_mf}
\end{figure*}

\begin{figure*}
    \centering
    \includegraphics[width=\textwidth]{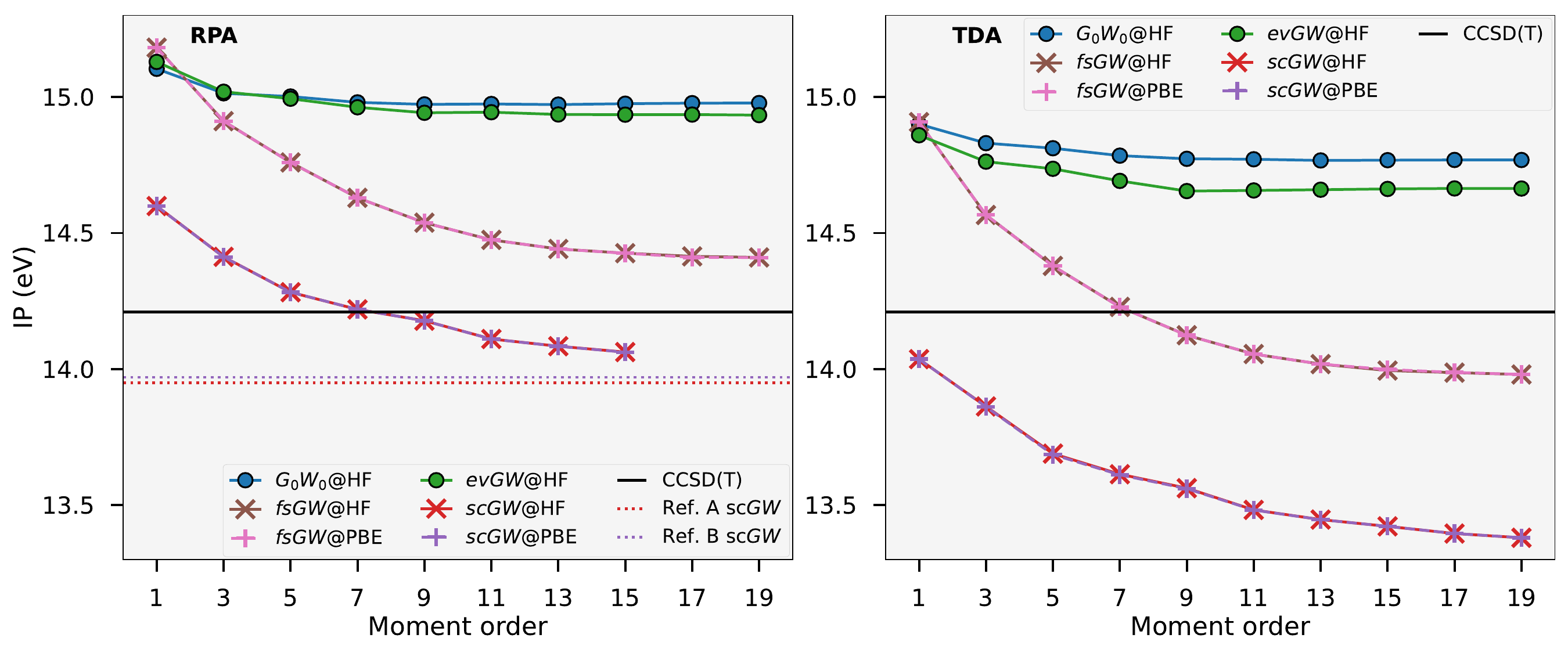}
    \caption{
        \toadd{Convergence of the IP of carbon monoxide (CO) using the def2-TZVPP basis set with respect to the number of conserved moments ($n_\textrm{mom}^\textrm{max}$) for various self-consistent implements across HF and PBE starting points, with RPA screening (left) and TDA (right). Reference fully self-consistent GW values are included from Caruso et al.~\cite{Caruso2016} (Ref. A) and Wen et al.~\cite{Wen2024} (Ref. B) where the convergence takes place on the Matsubara axis. The sc$GW$ results for RPA screening are limited to 15 moments due to convergence issues for higher moments.}
    }
    \label{fig:CO_gw}
\end{figure*}

\else
\begin{figure}
    \centering
    \includegraphics[width=\textwidth]{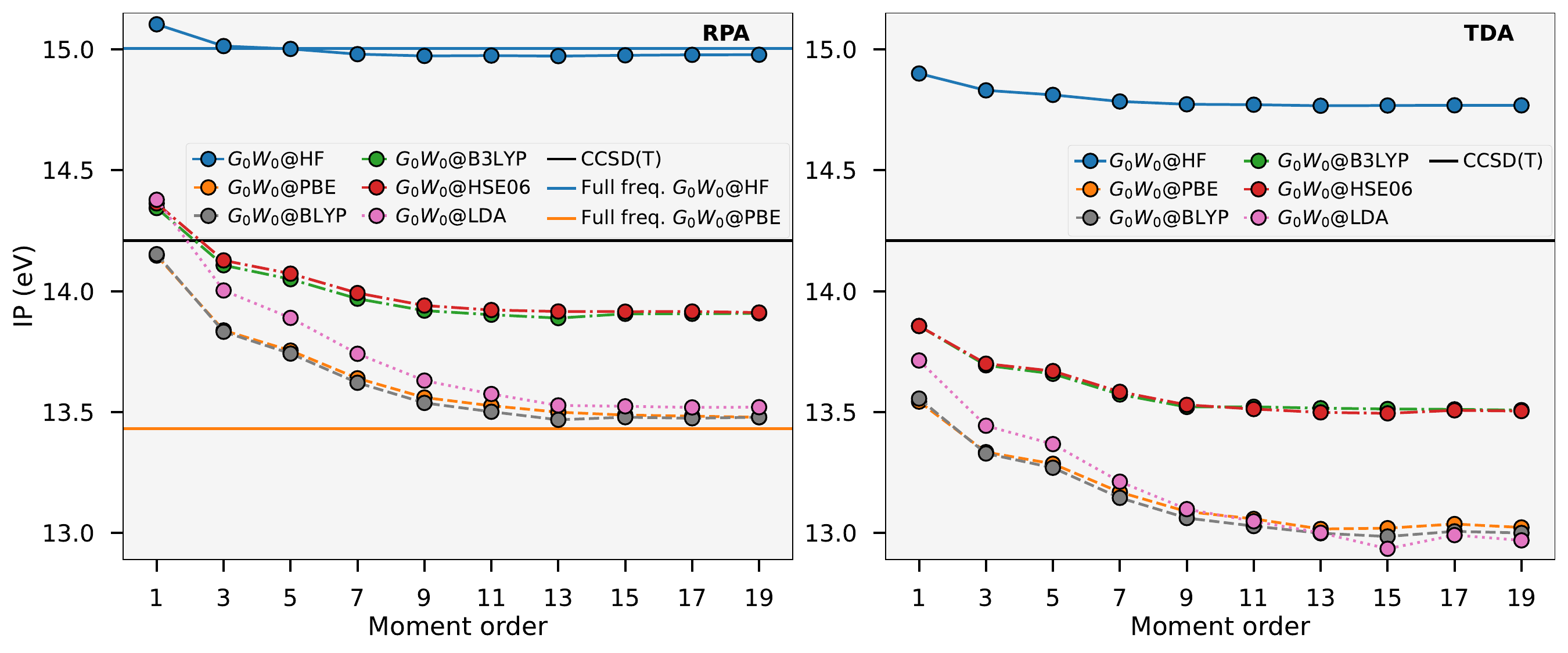}
    \caption{\toadd{Convergence of the IP of carbon monoxide ($CO$) with respect to the number of conserved moments ($n_\textrm{mom}^\textrm{max}$) of the self-energy in a def2-TZVPP basis set for single-shot $G_0W_0$, with RPA screening (left) and TDA (right). A range of mean-field starting points are considered, as well as reference values for RPA screening from {\tt PySCF}, implementing an $\mathcal{O}[N^6]$ full-frequency algorithm to remove any grid approximations \cite{doi:10.1021/acs.jctc.0c00704}. The remaining discrepancy likely comes from the diagonal approximation to the self-energy enforced in the reference values.}}
    \label{fig:CO_mf}
\end{figure}

\begin{figure}
    \centering
    \includegraphics[width=\textwidth]{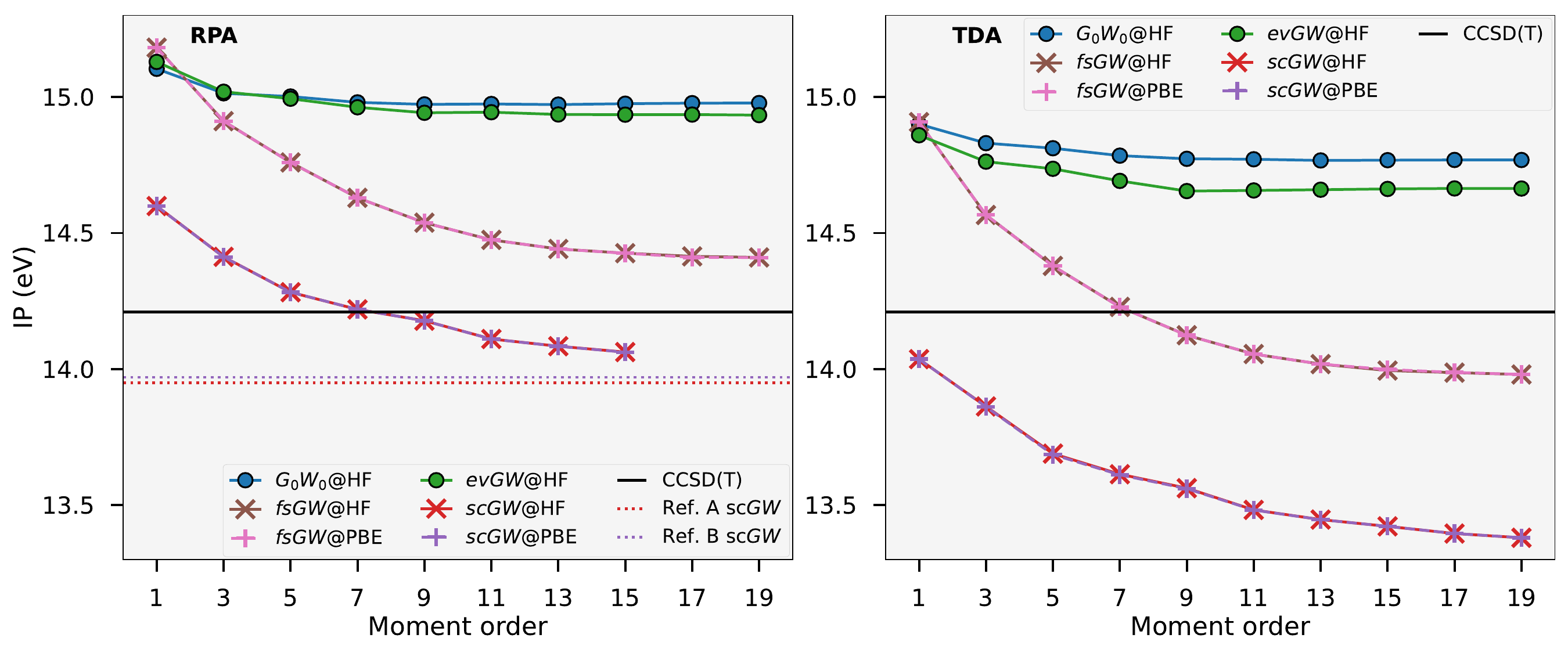}
    \caption{
        \toadd{Convergence of the IP of carbon monoxide ($CO$) using the def2-TZVPP basis set with respect to the number of conserved moments ($n_\textrm{mom}^\textrm{max}$) for various self-consistent implements across HF and PBE starting points, with RPA screening (left) and TDA (right). Reference fully self-consistent GW values are included from Caruso et al.~\cite{Caruso2016} (Ref. A) and Wen et al.~\cite{Wen2024} (Ref. B) where the convergence takes place on the Matsubara axis. The sc$GW$ results for RPA screening are limited to 15 moments due to convergence issues for higher moments.}
    }
    \label{fig:CO_gw}
\end{figure}
\fi






\end{document}